\documentclass[sigconf]{aamas}

\usepackage{balance} 

\usepackage{algorithm}
\usepackage{algorithmic}

\newtheorem{definition}{Definition}

\usepackage{multirow}
\usepackage{mathtools}
\usepackage{subcaption}
\usepackage{amsmath}
\usepackage{float}
\newcommand{\mysubsubsection}[1]{\noindent \textbf{#1}}

\newcommand{\Xv}{\boldsymbol{X}}
\newcommand{\xv}{\boldsymbol{x}}
\newcommand{\thetav}{\boldsymbol{\theta}}

\usepackage{cleveref} 

\setcopyright{ifaamas}
\acmConference[AAMAS '26]{Proc.\@ of the 25th International Conference
on Autonomous Agents and Multiagent Systems (AAMAS 2026)}{May 25 -- 29, 2026}
{Paphos, Cyprus}{C.~Amato, L.~Dennis, V.~Mascardi, J.~Thangarajah (eds.)}
\copyrightyear{2026}
\acmYear{2026}
\acmDOI{}
\acmPrice{}
\acmISBN{}

\acmSubmissionID{546}

\title{QD-MAPPER: A Quality Diversity Framework to Automatically Evaluate Multi-Agent Path Finding Algorithms in Diverse Maps}

\author{Cheng Qian}
\authornote{Equal contribution.}
\affiliation{
  \institution{Carnegie Mellon University}
  \city{Pittsburgh}
  \country{United States}
  }
\email{chengqia@andrew.cmu.edu}

\author{Yulun Zhang}
\authornotemark[1]
\affiliation{
  \institution{Carnegie Mellon University}
  \city{Pittsburgh}
  \country{United States}
  }
\email{yulunzhang@cmu.edu}

\author{Varun Bhatt}
\affiliation{
  \institution{University of Southern California}
  \city{Los Angeles}
  \country{United States}
  }
\email{vsbhatt@usc.edu}

\author{Matthew C. Fontaine}
\affiliation{
  \institution{University of Southern California}
  \city{Los Angeles}
  \country{United States}
  }
\email{mfontain@usc.edu}

\author{Stefanos Nikolaidis}
\affiliation{
  \institution{University of Southern California}
  \city{Los Angeles}
  \country{United States}
  }
\email{nikolaid@usc.edu}

\author{Jiaoyang Li}
\affiliation{
  \institution{Carnegie Mellon University}
  \city{Pittsburgh}
  \country{United States}
  }
\email{jiaoyangli@cmu.edu}

\begin{abstract}
We use the Quality Diversity (QD) algorithm with Neural Cellular Automata (NCA) to automatically evaluate Multi-Agent Path Finding (MAPF) algorithms by generating diverse maps. Previously, researchers typically evaluate MAPF algorithms on a set of specific, human-designed maps at their initial stage of algorithm design. However, such fixed maps may not cover all scenarios, and algorithms may overfit to the small set of maps. To seek further improvements, systematic evaluations on a diverse suite of maps are needed.
In this work, we propose \textbf{Q}uality-\textbf{D}iversity \textbf{M}ulti-\textbf{A}gent \textbf{P}ath Finding \textbf{P}erformance \textbf{E}valuato\textbf{R} (QD-MAPPER), a general framework that takes advantage of the QD algorithm to 
comprehensively understand the performance of MAPF algorithms by generating maps with patterns, be able to make fair comparisons between two MAPF algorithms, providing further information on the selection between two algorithms and on the design of the algorithms. Empirically, we employ this technique to evaluate and compare the behavior of different types of MAPF algorithms, including search-based, priority-based, rule-based, and learning-based algorithms. Through both single-algorithm experiments and comparisons between algorithms, researchers can identify patterns that each MAPF algorithm excels and detect disparities in runtime or success rates between different algorithms. Our generated maps are available at \url{https://airtclick.github.io/qdmapper}. Our code is available at \url{https://github.com/AirTClick/QD-MAPPER}.
\end{abstract}


\newcommand{\BibTeX}{\rm B\kern-.05em{\sc i\kern-.025em b}\kern-.08em\TeX}

\begin{document}


\pagestyle{fancy}
\fancyhead{}


\maketitle 


\section{Introduction}

We study the problem of evaluating the performance of Multi-Agent Path Finding (MAPF) algorithms by generating diverse maps. Given a map and a group of agents, MAPF is the problem of finding collision-free paths from their start to goal locations. MAPF has wide applications in coordinating hundreds of robots in automated warehouses~\cite{ZhangNCA2023,zhangLayout23}, 
moving characters in video games~\cite{MaAIIDE17}, 
managing drone traffic~\cite{choudhury2022TruckUAVmapf}, and controlling multi-robotic arms~\cite{ShaoulICAPS24}.

Given the wide applicability of MAPF, many algorithms have been proposed to solve MAPF, and evaluating these algorithms becomes an important task. 
At the beginning of the algorithm design, researchers prefer to use benchmark maps to have an overview on the performance of their algorithms. 
\citet{SternSoCS19} have proposed a set of 33 MAPF benchmark maps,
covering a diverse spectrum of map sizes, layouts, and difficulties. These human-designed maps are widely used to test newly proposed MAPF algorithms~\cite{Li2020EECBSAB,okumura2019priority,learntofollow2024}. However, to further discover the pros and cons and improve algorithms, such fixed maps are not sufficient.

Meanwhile, Quality Diversity (QD) algorithms have been used to generate a diverse set of high-quality solutions by optimizing a given objective function and a set of diversity measure functions. A recent work~\cite{zhangLayout23} has used QD algorithms to optimize layouts for automated warehouses. To ensure that the optimized layouts possess human-explainable and regularized patterns, a follow-up work~\cite{ZhangNCA2023} then leverages Neural Cellular Automata (NCA), a CNN-based neural generator, to generate the layouts and uses QD algorithms to optimize the parameters of the NCA. Cellular Automata~\cite{gardner1970} iteratively generates complex cell-based structures from a simple one through local interaction between cells. Each cell decides its next state based on its neighbors via a fixed rule. NCA then represents the rules using a convolutional neural network that can be optimized.


In this paper, we adapt the layout optimization approach from the previous work~\cite{ZhangNCA2023}, proposing \textbf{Q}uality-\textbf{D}iversity \textbf{M}ulti-\textbf{A}gent \textbf{P}ath Finding \textbf{P}erformance \textbf{E}valuato\textbf{R} (QD-MAPPER), a general framework with the goal of evaluating MAPF algorithms by generating diverse maps. To demonstrate that our approach can evaluate a broad spectrum of MAPF algorithms, we present evaluation results on 6 representative MAPF algorithms, namely CBS~\cite{SharonAIJ15}, EECBS~\cite{Li2020EECBSAB}, PBS~\cite{MaAAAI19}, LaCAM3~\cite{okumura2024lacam3}, PIBT~\cite{okumura2019priority}, and Learn-to-Follow (LTF)~\cite{learntofollow2024}. We present a number of interesting results undiscovered in prior works. For example, we generate new maps with patterns that LaCAM3 runs out of time, while it was reported to solve 99\% of the MAPF benchmarks~\cite{SternSoCS19}. We also expose the incompleteness of PBS by presenting maps in which PBS fails to find any solution instead of running out of time on benchmark maps. 


We make the following contributions: 
(1) we adapt the layout optimization approach to propose QD-MAPPER to evaluate MAPF algorithms by generating diverse maps,
(2) we apply QD-MAPPER on 6 representative MAPF algorithms of different categories, finding new challenging maps for MAPF algorithms, and providing new insights about their performance on maps of different patterns, and
(3) we propose two concrete realizations of QD-MAPPER that are useful for evaluating MAPF algorithms: \emph{one-algorithm} realization and \emph{two-algorithm} realization. As a generic framework, researchers can extend other realizations, such as generating easy maps or comparing more than two algorithms.

\section{Preliminaries}

\subsection{Multi-Agent Path Finding (MAPF)} \label{subsec:MAPF}

\begin{definition}[Map]
A map is a four-neighbored 2D grid, where each tile can either be an empty space or an obstacle. A map is \emph{valid} if all empty spaces are connected.
\end{definition}

\begin{definition}[MAPF]
Given a valid map and a set of agents with their start and goal locations, MAPF aims to find collision-free paths from their start to goal locations. Agents can either move to their adjacent locations or stay at their current locations at each timestep. Two agents collide if they are at the same location or swap locations at the same timestep. The objective of MAPF is minimizing the sum-of-cost, defined as the sum of travel time of all agents.
\end{definition}

We present a short summary of existing MAPF algorithms and the chosen algorithms for experiments in this paper.


\mysubsubsection{Search-Based.} This category includes exponential-time algorithms that exhaustively explore the solution space of MAPF. They usually have theoretical guarantees such as optimality and bounded suboptimality, but suffer from long computational times. 
Examples include M$^*$~\cite{MStar}, 
BCP~\cite{LamIJCAI19}, 
ICTS~\cite{SharonAIJ13}, 
CBS~\cite{SharonAIJ15}, 
ECBS~\cite{BarrerSoCS14}, 
and EECBS~\cite{Li2020EECBSAB}
. 
We choose CBS~\cite{SharonAIJ15} and EECBS~\cite{Li2020EECBSAB}
as the representatives of this category.
CBS is an optimal algorithm. It starts by planning a shortest path for each agent, ignoring the collisions, and then resolving the collisions with a two-level search. 
The high-level search iteratively selects unresolved collisions and adds constraints to tackle them. The low-level search runs single-agent planning to compute new paths that satisfy the constraints. The search continues until all collisions are resolved. 
EECBS is a bounded-suboptimal variant of CBS that uses Explicit Estimation Search~\cite{Thayer2011EES} on the high level and focal search~\cite{PearTPAMIl1982} on the low level. 

\mysubsubsection{Priority-Based.} Priority-based algorithms plan paths for each agent following a priority order, forcing agents with lower priorities to avoid colliding with those with higher ones. 
The priority planning (PP)~\cite{Erdmann87,StandleyAAAI10} algorithm plans paths for agents in a pre-defined priority order. Monte-Carlo PP~\cite{friedrich2024scalable} pre-defines a set of randomized priority orders and selects the one with the best paths. 
Representative algorithms include priority planning (PP)~\cite{Erdmann87}, Monte-Carlo PP~\cite{friedrich2024scalable}, and PBS~\cite{MaAAAI19}.
We choose PBS as the representative of this category.
PBS combines CBS with PP to explicitly search for a good priority order with a two-level search. The high-level search of PBS is similar to CBS except that PBS constrains one agent to have a higher priority order than the other. The low-level search then plans paths with the order from the high-level search.

\mysubsubsection{Rule-Based.}
Rule-based algorithms \cite{WangB11} leverage pre-defined rules to move agents to their goals. They usually run much faster, but produce worse solutions than search-based and priority-based methods.
We choose PIBT~\cite{okumura2019priority} and LaCAM3~\cite{okumura2024lacam3} as representatives.
PIBT uses an iterative one-timestep rule to move agents. At each timestep, each agent plans a single-step action towards its goal. In case of collisions, a pre-defined rule is applied to resolve the collision. LaCAM3 is an anytime algorithm that combines PIBT and search-based methods.
As a hybrid algorithm that combines PIBT with search-based methods, It performs the search by sequentially expanding search nodes, and when it encounters the goal configuration, it derives the solution by backtracking parent pointers.

\mysubsubsection{Learning-Based.}
Learning-based algorithms either formulate MAPF as a Multi-Agent Reinforcement Learning (MARL)~\cite{guillaume2018primal} problem or leverage Imitation Learning (IL) to imitate the behavior of an oracle MAPF algorithm. MARL-based methods \cite{guillaume2018primal,Wang2023SCRIMPSC,learntofollow2024} typically train a shared policy for each agent, taking its local field of view as input and deciding its next action. IL-based methods, on the other hand, attempt to imitate an oracle algorithm. For example, MAPF-GPT~\cite{Andreychuk_MAPFGPT_2025} tokenizes MAPF instances and trains a transformer to imitate LaCAM3~\cite{okumura2024lacam3}. SILLM~\cite{JiangICRA25} trains a learnable PIBT by imitating the behavior of WPPL~\cite{Jiang2024Competition}.
We choose Learn to Follow (LTF)~\cite{learntofollow2024} as the representative of the learning-based algorithms.
LTF starts by searching for a guide path for each agent without considering collisions. 
Then it uses a shared learned policy to move the agents to their goals along the guide paths while avoiding collisions. LTF is developed for lifelong MAPF, a variant of MAPF that constantly assigns new goals to agents. We modify it for MAPF by asking agents to stop when they reach their first goals.

\subsection{MAPF Algorithm Evaluation}


In early works, researchers typically used randomly generated benchmark maps~\cite{MStar} or maps from the single-agent path planning benchmark set~\cite{sturtevant2012benchmarks} to evaluate new MAPF algorithms. 
\citet{SternSoCS19} established the major set of MAPF benchmark maps with 33 maps, which are used extensively in MAPF research~\cite{Li2020EECBSAB,Chan2023GreedyPS,zhang2024ggo}.
These maps can provide an overview of the performance of MAPF algorithms at the initial stage. However, to further discover the pros and cons of the algorithm, the benchmark set is insufficient with its fixed patterns and limited size. For example, LaCAM3, despite prior claims
of solving 99\% of standard benchmarks, also struggles under our generated maps with long corridors and narrow entries, which are patterns underrepresented in existing benchmarks.

To the best of our knowledge, very few works use automatic generation methods to evaluate the performance of MAPF algorithms. The POGEMA platform~\cite{skrynnik2025pogema} leverages procedural content generation~\cite{Cobbe2019LeveragingPG} to train and evaluate MARL agents, but it does not target specific MAPF algorithms.
One recent work~\cite{ren2024MAPFhard} also uses QD to generate maps for MAPF. However, it has two major differences with our work. First, they focus on generating maps of fixed difficulties, quantified by an approximate metric based on map connectivity, while we are interested in evaluating the performance of specific MAPF algorithms by generating diverse maps with different difficulties. Second, they directly optimize tile types (i.e., whether it is an empty space or an obstacle), while we optimize a map generator based on NCA, which has been shown to be more effective for generating maps with diverse regularized patterns~\cite{ZhangNCA2023}.




\subsection{Quality Diversity (QD) Algorithms and Automatic Scenario Generation}
QD algorithms~\cite{mouret2015illuminating} are inspired by evolutionary algorithms with diversity search to generate a diverse collection of high-quality solutions by optimizing an objective function and diversifying a set of diversity measure functions. QD algorithms maintain an \emph{archive}, which is a tessellated measure space defined by the measure functions. The archive stores the best solution in each tessellated cell. QD algorithms then optimize the sum of objective values of all solutions in the archive, defined as QD-score.
We choose Covariance Matrix Adaptation MAP-Annealing (CMA-MAE)~\cite{Fontaine2022CovarianceMA} as the method to generate maps because it is the state-of-the-art QD algorithm specialized for continuous search domains. 
CMA-MAE is an extension of MAP-Elites~\cite{mouret2015illuminating} that incorporates the covariance matrix adaptation mechanism of CMA-ES~\cite{hansen2016cmaes}, which is a derivative-free single-objective optimizer. CMA-ES maintains a multi-variate Gaussian distribution and iteratively samples from it for new solutions. It evaluates the solutions and updates the Gaussian toward high-objective regions. CMA-MAE adapts this mechanism to optimize the QD-score.
QD algorithms are widely applied in automatic scenario generation scenarios.
In multi-robot systems, ~\citet{ZhangNCA2023} optimize NCA via QD algorithms to generate arbitrarily large warehouse layouts.
In autonomous driving, researchers generate scenarios to evaluate developed autonomous driving systems~\cite{Abeysirigoonawardena2019GeneratingAD,Mullins2018AdaptiveGO}. In human-robot interaction, QD algorithms are used to generate shared autonomy scenarios~\cite{fontaine2021quality} and diverse kitchen layouts~\cite{fontaine2021importance} to study the coordination behavior between humans and robots.
A follow-up work~\cite{Bhatt2023surrHRI} leverages model-based QD methods~\cite{Bhatt2022DeepSA,Zhang2021DeepSA} to improve the sample efficiency of QD algorithms. 
In reinforcement learning, prior works have generated maps to benchmark~\cite{Cobbe2019LeveragingPG} 
or continuously improve~\cite{Wang2019PairedOT} 
trained agents.

\section{Map Generation Approach}

\begin{figure}[!t]
    \centering
    \includegraphics[width=0.47\textwidth]{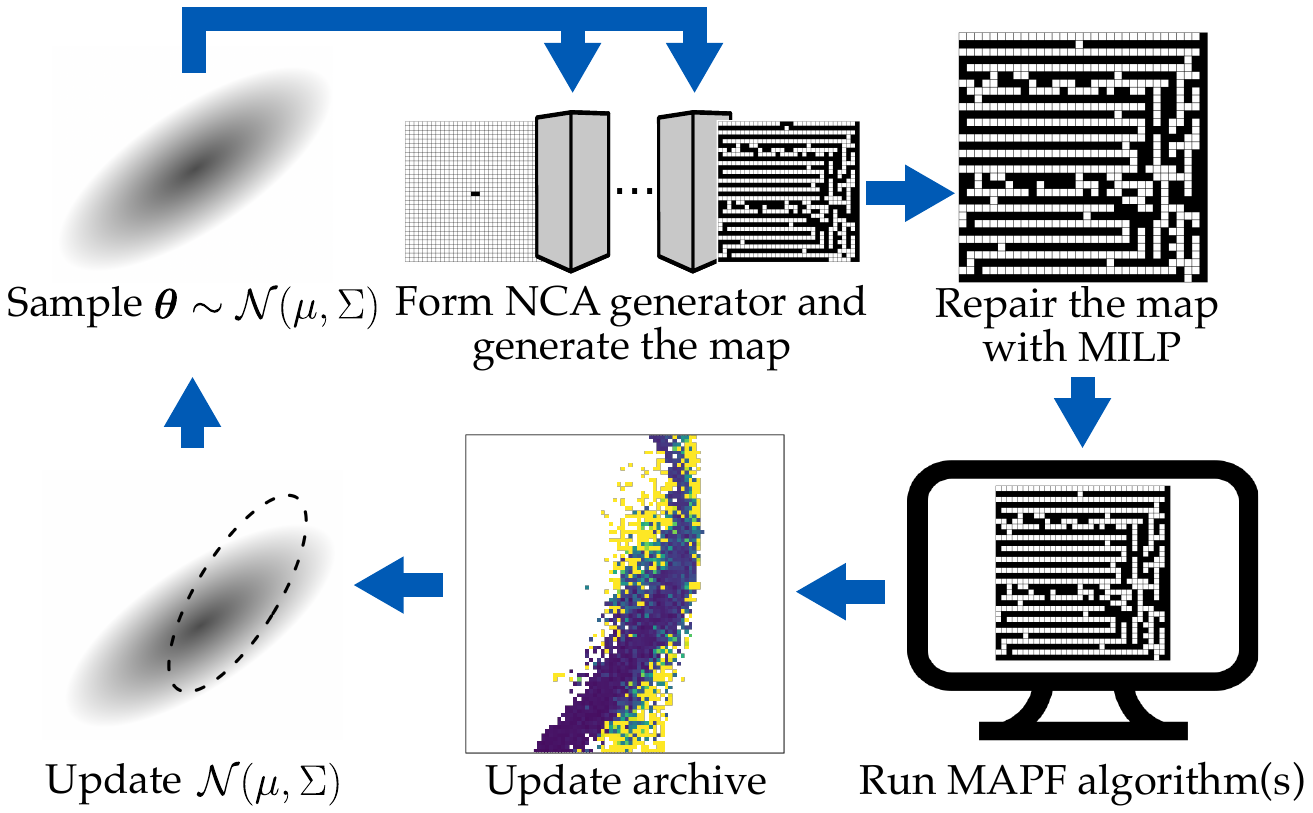}
    \caption{Overview of QD-MAPPER:
    (1) sample parameter vectors $\thetav$ from multi-variate Gaussian of CMA-MAE, 
    (2) form NCA generators,
    (3) use a Mixed Integer Linear Programming (MILP) solver to ensure map connectivity and the correct number of obstacles,
    (4) run MAPF algorithms, 
    (5) update the archive, and
    (6) update multi-variate Gaussian.}
    \Description[Overview of QD-MAPPER]{The pipeline shows how QD-MAPPER generates maps to evaluate the performance of targeted algorithms.}
    \label{fig:pipe}
\end{figure}




\subsection{Overview}

\Cref{fig:pipe} shows the pipeline of QD-MAPPER.
We adapt previous works to use CMA-MAE~\cite{Fontaine2022CovarianceMA} to search for and update a diverse collection of NCA~\cite{ZhangNCA2023} generators to generate diverse maps with the objective and measures computed by running MAPF algorithms. 
We start by sampling a batch of $b$ parameter vectors $\theta$ from a multivariate Gaussian distribution, forming $b$ NCA generators. 
Starting from a fixed seed map, each NCA generator iteratively updates the input map to generate a map with complex local patterns, leading to $b$ maps. 
\Cref{appen:nca} exemplifies the process of generating a map.   

Maps generated by NCA might not be valid. 
We then adapt a Mixed Integer Linear Programming (MILP) solver~\cite{zhang:aiide2020,zhangLayout23} to repair the map to enforce connectivity and domain-specific constraints while making minimal modifications to the generated map. 
We then evaluate repaired maps with MAPF algorithms to calculate the average objectives and measure values, and add the evaluated maps to an archive. 
For each evaluation, we run the given MAPF algorithm in $N_e$ instances. 
Finally, we update the parameters of the multivariate Gaussian distribution, sampling a new batch of $b$ parameter vectors and starting a new iteration. We repeat this process until we evaluate $N_{eval}$ maps.



\mysubsubsection{NCA generator.}
We define an NCA generator as a function $g(s; \theta; C) : M \to M$, where $M$ is the space of possible maps, \( s \in M \) represents a fixed seed map, and \( C \in \mathbb{Z}^{+} \) is a hyperparameter controlling the number of iterations to run the NCA generator. The generator is parameterized by \( \theta \in \Theta \), where each parameter vector \( \theta \) corresponds to a distinct NCA generator that then defines a distinct map \( m \in M \). We then use CMA-MAE to optimize $\theta$, attempting to find the best NCA generator in each cell of the archive.

Our NCA generator is a convolutional neural network (CNN) following the architecture in prior work ~\cite{ZhangNCA2023}. It consists of three convolutional layers with kernel size $3 \times 3$, each followed by either ReLU or sigmoid activations. We represent the 2D tile-based maps as 3D tensors, with each tile encoded as a one-hot vector, representing its tile type (i.e. obstacle or empty space). Since the output of the NCA has the same size as its input, we can repeatedly pass the output back into the network. 


\mysubsubsection{MILP Solver.}
Maps generated by NCA might not be valid (e.g., a wall partitioning two open sections).
To ensure that the maps are valid, we formulate and solve a MILP~\cite{zhang:aiide2020,zhangLayout23} to obtain a repaired map. 
The MILP constraints are set to be: (1) all empty spaces are connected, and (2) the number of obstacles falls within a pre-defined range $[O_{lb}, O_{ub}]$.
We set the objective to minimize the Hamming distance between unrepaired and repaired maps to ensure that the patterns of the generated map are maintained as much as possible. 


\mysubsubsection{Generation of MAPF Instances From Maps.}
To generate a MAPF \emph{instance} based on a map,
we follow the bucket method~\cite{SternSoCS19} to generate evenly distributed start and goal locations of the agents with a distance constraint, which was used for generating the \emph{even} scenarios in the MAPF benchmark. For each map, we generate $N_e$ MAPF instances with different starts and goals, run MAPF algorithms on each instance, and compute the average objective and measures of all instances as the objective and measures of the map. 
Notably, our MAPF instance generation strategy serves only as an example. 
In applications where a different set of instances is needed, users can customize the instance generation method inside QD-MAPPER.




\mysubsubsection{Framework Realization.}
Our proposed framework is flexible at a high level. 
In this paper, we propose two concrete realizations of QD-MAPPER that are useful for evaluating MAPF algorithms with carefully designed objectives and measures. We first provide a \emph{one-algorithm} realization, intending to evaluate five representative MAPF algorithms by generating diverse and difficult maps for each of them. We then provide a \emph{two-algorithm} realization, aiming to compare the performance of two algorithms by generating maps that are easy for one algorithm and hard for the other. We highlight that other realizations, such as generating easy maps or comparing more than two algorithms, are possible.




\subsection{Objective}
In all experiments, our objective is a function $f: \Xv \rightarrow \mathbb{R}$, where $\Xv$ is the space of all maps. The function $f$ runs one or two MAPF algorithms on $N_e$ different MAPF instances with $N_a$ agents. 
For CBS, EECBS, LaCAM3, and PBS, we set a time limit $T$ for each run. For PIBT and LTF, we set a maximum makespan of $M$ timesteps. 

\mysubsubsection{One-Algorithm Experiments.}
We intend to generate maps that are challenging for a given MAPF algorithm $\phi \in A$, where $A = \{\text{CBS},\ \text{EECBS},\ \text{LaCAM3},\ \text{PBS},\ \text{PIBT},\ \text{LTF}\}$ in our experiments. The objective quantifies the empirical hardness of a map $\xv \in \Xv$ to $\phi$.

For CBS, EECBS, and PBS, the objective is to maximize their \emph{CPU runtime} because they are exponential-time algorithms, the major limitation being frequently running out of time. We do not consider the solution quality as the objective because CBS and EECBS are proven to be optimal and bounded-suboptimal algorithms. 
PBS, while being unbounded suboptimal, empirically finds near-optimal solutions. For example, \citet{MaAAAI19} shows that PBS never finds solutions over 4\% worse than optimal on game maps from the MAPF benchmark.
We include LaCAM3 because it can solve almost all instances in the MAPF benchmark. Our objective is to identify more challenging maps for it.
Therefore, we maximize \emph{CPU runtime} to find failure cases for CBS, EECBS, PBS, and LaCAM3. Concretely, suppose that the function $t_{\phi}: \Xv \rightarrow \mathbb{R}_{>0}$ computes the average CPU runtime by running $\phi \in A$ on $N_e$ instances, the objective $f$ for the algorithm $\phi$ is $f(\xv) = t_{\phi}(\xv), \phi \in \{\text{CBS},\ \text{EECBS},\ \text{PBS},\ \text{LaCAM3}\}$. If $\phi$ cannot find solutions within the time limit $T$, we set the CPU runtime to $T$.

PIBT and LTF, on the other hand, suffer mainly from deadlocks, meaning that some agents might never reach their goals.
Therefore, to find failure cases, we minimize the \emph{Regularized Success Rate (RSR)}, defined as follows:
\begin{equation}
    RSR_{\phi}(\xv) =
    \begin{cases}
        SR_{\phi}(\xv) & \text{if } SR < 1\\
        SR_{\phi}(\xv) \cdot C - SoC_{\phi}(\xv) & \text{if } SR = 1
    \end{cases}
\end{equation}
$SR_{\phi}: \Xv \rightarrow \mathbb{R}_{>0}$ computes the success rate of running algorithm $\phi$ on one MAPF instance, which is computed as the percentage of agents that successfully reach their goals within the given makespan $M$. $SoC_{\phi}: \Xv \rightarrow \mathbb{R}_{>0}$ computes the sum-of-cost of the solution if $SR = 1$. $C$ is a large constant making sure that the RSR score of maps with $SR = 1$ always dominates that of maps with $SR < 1$.
Intuitively, if not all agents reach their goals ($SR < 1$), we maximize the success rate. If all agents reach their goals ($SR = 1$), we regularize the success rate with the sum-of-cost to quantify the quality of the solution. Let the function $r_{\phi}:\Xv \rightarrow \mathbb{R}_{>0}$ return the average regularized success rate by running algorithm $\phi$ in $N_e$ MAPF instances. Concretely, $r_{\phi}(\xv) = \sum_{i=1}^{N_e} RSR_{\phi}^{(i)}(\xv)$, where $i$ denotes the $i$-th run of algorithm $\phi$. Then the objective of PIBT and LTF are $f(\xv) = r_{\text{PIBT}}(\xv)$ and $f(\xv) = r_{\text{LTF}}(\xv)$, respectively. We do not consider runtime as the objective of PIBT and LTF because they are both one-step planners and thus their runtime is strongly correlated with success rates and solution quality.

\mysubsubsection{Two-Algorithm Experiments.} To compare two given MAPF algorithms, we aim to generate maps that maximize the performance gap between the them. We consider two pairs of comparisons: (1) EECBS vs. PBS and (2) PIBT vs. LTF. We compare EECBS and PBS because (1) both suboptimal algorithms use a two-level search based on collision avoidance, (2) both find near-optimal solutions empirically, and (3) few prior works have compared them thoroughly. 
We compare PIBT and LTF because (1) both use a pre-defined policy, either rule-based or learned, to move the agents step by step to their goals while avoiding collisions, (2) both run fast but suffer from deadlocks, and (3) while the LTF work~\cite{learntofollow2024} has compared them on warehouse maps, showing that LTF outperforms PIBT, we want to find cases where PIBT outperforms LTF.
%
To compare EECBS and PBS, we set the objective as the absolute difference in the average CPU runtime. Concretely, the objective of a generated map $\xv \in \Xv$ is computed as $f(\xv) = |t_{\text{EECBS}}(\xv) - t_{\text{PBS}}(\xv)|$. Similarly, to compare PIBT and LTF, we set the objective as the absolute difference in the regularized success rate, i.e., $f(\xv) = |r_{\text{PIBT}}(\xv) - r_{\text{LTF}}(\xv)|$.

\subsection{Diversity Measures}
We select diversity measures such that the generated maps are diverse in (1) general hardness towards all algorithms to analyze whether the general hardness of a map aligns with the hardness to a specific algorithm, and (2) spatial arrangement of the obstacles to generate maps of different patterns.

\mysubsubsection{General Hardness.} 
To measure the general hardness of a map, the most commonly used metric is the number of obstacles.
The number of obstacles in a map affects the space available to resolve collisions as well as the search space of the algorithms. Therefore, diversifying it would give a range of difficulties.
We also consider two metrics that are proposed specifically to reflect the hardness of a map: the standard deviation of Betweenness Centrality (Std of BC)~\cite{ewing2022BC} and $\lambda_2$~\cite{ren2024MAPFhard}. 

The Betweenness Centrality of an empty space is the fraction of all possible shortest paths in the map that pass through it. We use the Std of BC here to measure the congestion in the map.
The second smallest eigenvalue (known as $\lambda_2$) of the normalized Laplacian matrix of different maps is also claimed to be correlated to the difficulty of the maps~\cite{ren2024MAPFhard}. 

\mysubsubsection{Spatial Arrangement.} In addition to the hardness of the maps, we want to diversify the spatial arrangement of the maps, making the generated maps stylistically diverse. We first consider the KL divergence of the tile pattern distribution~\cite{fontaine2020illuminating} between the tile pattern distribution of the generated map and a set of maps selected from the MAPF benchmark~\cite{SternSoCS19}. 

A \emph{tile pattern} is one possible arrangement of obstacles and empty spaces in a 3 $\times$ 3 grid. To compute the tile pattern distribution of a map, we count the number of possible tile patterns in the map. We then compute the KL divergence between the tile pattern distribution of the map and a fixed distribution calculated from all maze maps\footnote{\url{https://movingai.com/benchmarks/grids.html}} of the single-agent path finding benchmark~\cite{sturtevant2012benchmarks}. 
The smaller the KL divergence, the more similar in local patterns our generated map is to the selected set of benchmark maps. 
We choose maze maps to compare the KL divergence with because they empirically yield the most complex patterns. 

In addition, we consider the entropy of the tile pattern distribution~\cite{ZhangNCA2023}. The entropy of the tile pattern distribution is a diversity measure to indicate the level of pattern regularization. 
The lower the entropy, the more regularized patterns the map possesses.

Finally, we consider the KL divergence of the Weisfeiler-Lehman (WL) graph feature~\cite{Nino2011WLkernal} between the generated map and the maze maps. The WL graph feature indicates the similarity between graphs. 
Similar to the tile pattern distribution, a smaller KL divergence of the WL graph feature implies more similarities between the local patterns of the generated map and the maze maps.

In our experiments, we choose \textbf{the number of obstacles} and \textbf{KL divergence of tile pattern distribution} as the diversity measures. We empirically observe that, among the measures considered, the number of obstacles is the most correlated with the general hardness of the maps, and the KL divergence of tile pattern distribution yields the most diverse map patterns. We justify our choice with experimental results in \Cref{appen:measures}.

\section{One Algorithm Analysis} \label{sec:one-algo}

For one-algorithm experiments, we intend to generate diverse difficult maps for CBS, EECBS, PBS, LaCAM3, PIBT, and LTF. 


\begin{figure}[!t]
    \centering
    \begin{subfigure}[b]{0.14\textwidth}
      \centering
      \includegraphics[width=1\textwidth]{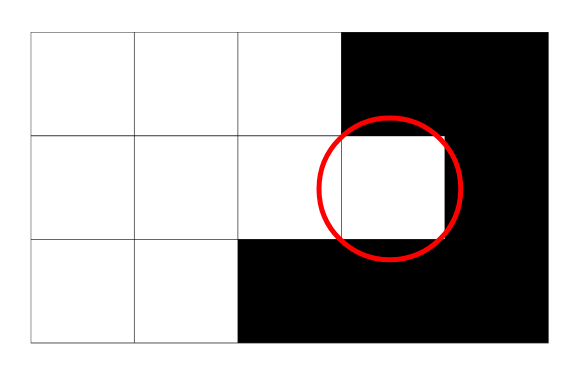}
      \caption{One-entry space.}
      \label{fig:one-entry-space}
    \end{subfigure}%
    \hspace{0.03\textwidth}
    \begin{subfigure}[b]{0.14\textwidth}
      \centering
      \includegraphics[width=1\textwidth]{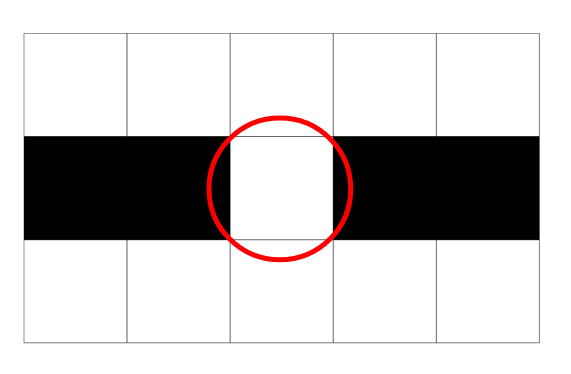}
      \caption{One-tile entry.}
      \label{fig:one-tile-entry}
    \end{subfigure}
    \caption{
    Examples of one-entry space and one-tile entry.
    }
    \Description[Examples of one-entry space and one-tile entry.]{One-entry space and one-tile entry are two major patterns that commonly appear in generated maps.}
    \label{fig:example-bad}
\end{figure}

\begin{figure}[!t]
    \centering
    \includegraphics[width=0.5\textwidth]{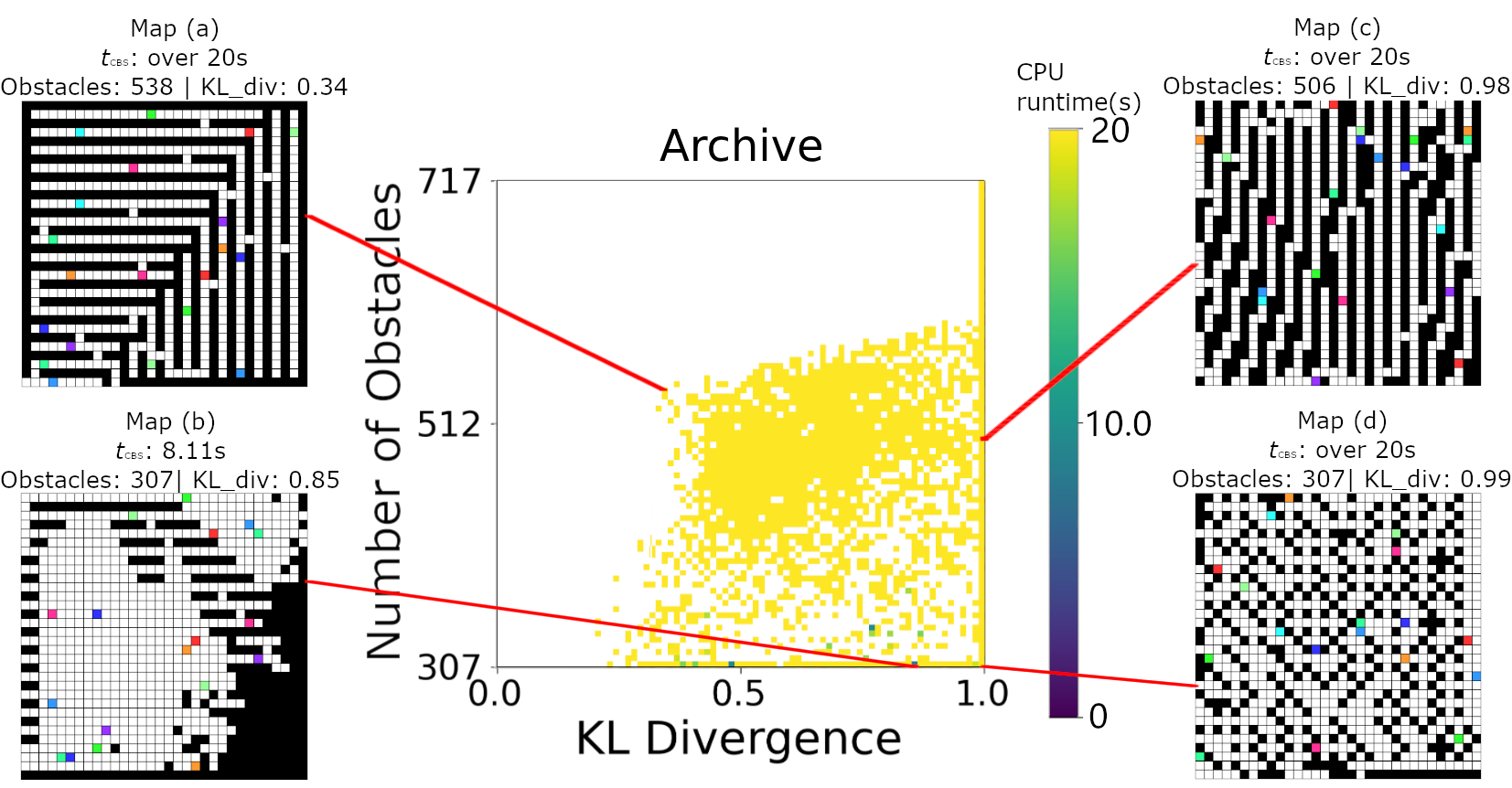}
    \caption{Archive for CBS with sample maps.}
    \Description[Archive of CBS]{Archive of CBS shows the distribution of difficult maps QD-MAPPER generates with four representative maps.}
    \label{fig:CBS}
\end{figure}

\subsection{Experiment Setup}
\mysubsubsection{Hyperparameters.} \label{subsec:one-algo-exp-setup}
Defining $w$ as the suboptimality bound away from the optimal solution, we use $w = 1.5$ for EECBS.
For CBS, we use the implementation of EECBS with $w = 1$. We set $T = 20$ seconds for CBS, EECBS, LaCAM3, and PBS, $M = 1000$ for PIBT, and $M = 512$ for LTF.
For all algorithms, we fix the map size as 32 $\times$ 32.
The range for the number of obstacles is $[O_{lb}=307, O_{ub}=717]$, which corresponds to 30\% to 70\% of the map size.
For each evaluation, we run the given MAPF algorithm in $N_{e} = 5$ different instances, each with $N_a$ agents. For CBS, EECBS, and PBS, $N_{a} = 50$, for LaCAM3, $N_{a} = 100$, and for PIBT and LTF, $N_{a} = 150$. We use more agents for LaCAM3, PIBT, and LTF because they empirically have higher success rates in congested MAPF instances. 
For all algorithms, we evaluated $N_{eval} = 10,000$ maps. 
We present computational costs and resources in \Cref{appen:implement-compute}.

\mysubsubsection{Implementation.}
We implement CMA-MAE with Pyribs~\cite{pyribs}. 
For MAPF algorithms, 
we use the open source implementations of CBS and EECBS\footnote{\url{https://github.com/Jiaoyang-Li/EECBS}},
PBS\footnote{\url{https://github.com/Jiaoyang-Li/PBS}},
LaCAM3\footnote{\url{https://github.com/Kei18/lacam3}},
PIBT\footnote{\url{https://github.com/Kei18/pibt2}}, and
LTF\footnote{\url{https://github.com/Cognitive-AI-Systems/learn-to-follow}}.


\subsection{Results} 

\begin{figure}[!t]
    \centering
    \includegraphics[width=0.5\textwidth]{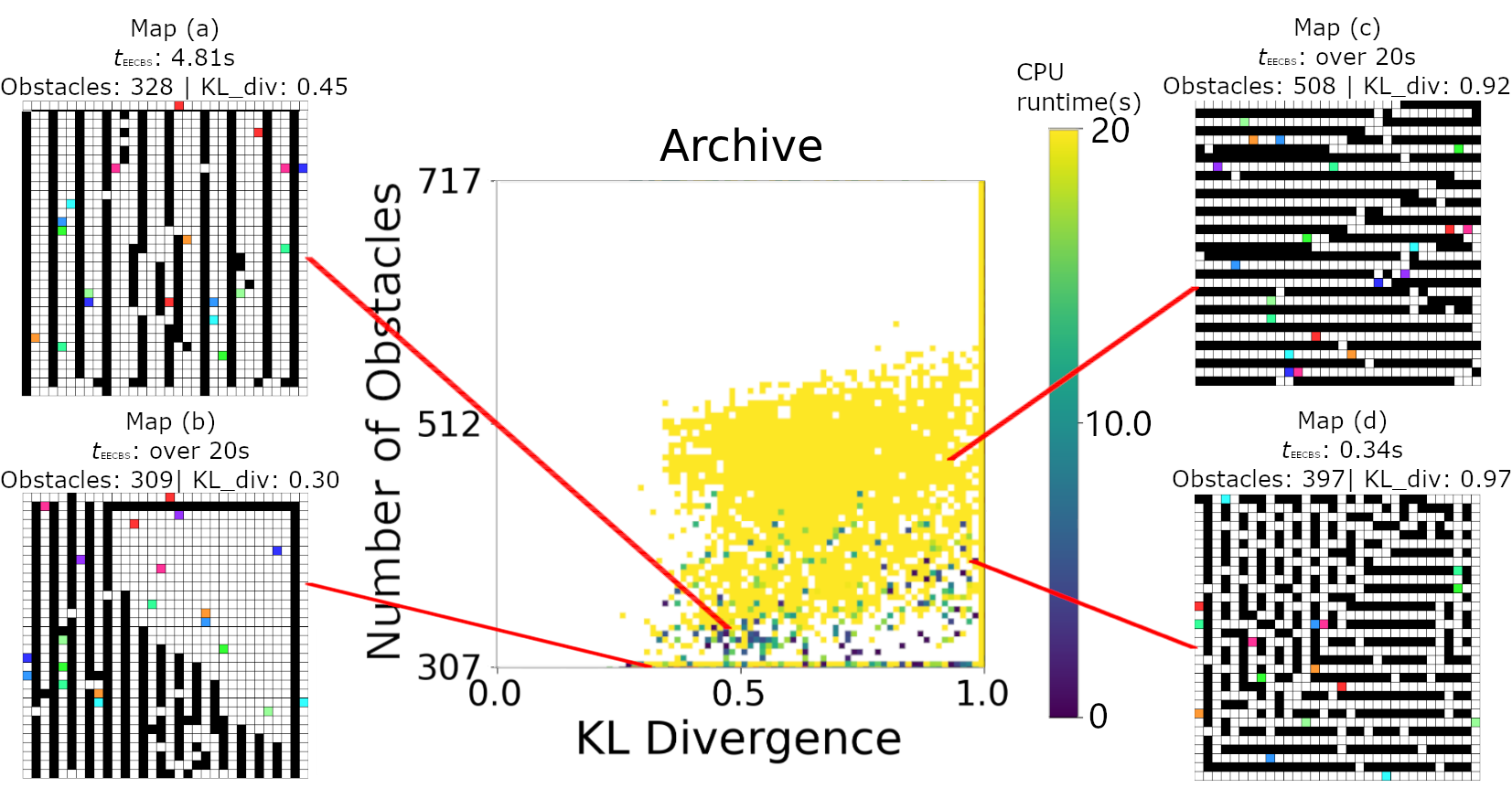}
    \caption{Archive for EECBS with sample maps.
}
    \Description[Archive of EECBS]{Archive of EECBS shows the distribution of difficult maps QD-MAPPER generates with four representative maps.}
    \label{fig:EECBS}
\end{figure}

\begin{figure}[!t]
    \centering
    \includegraphics[width=0.5\textwidth]{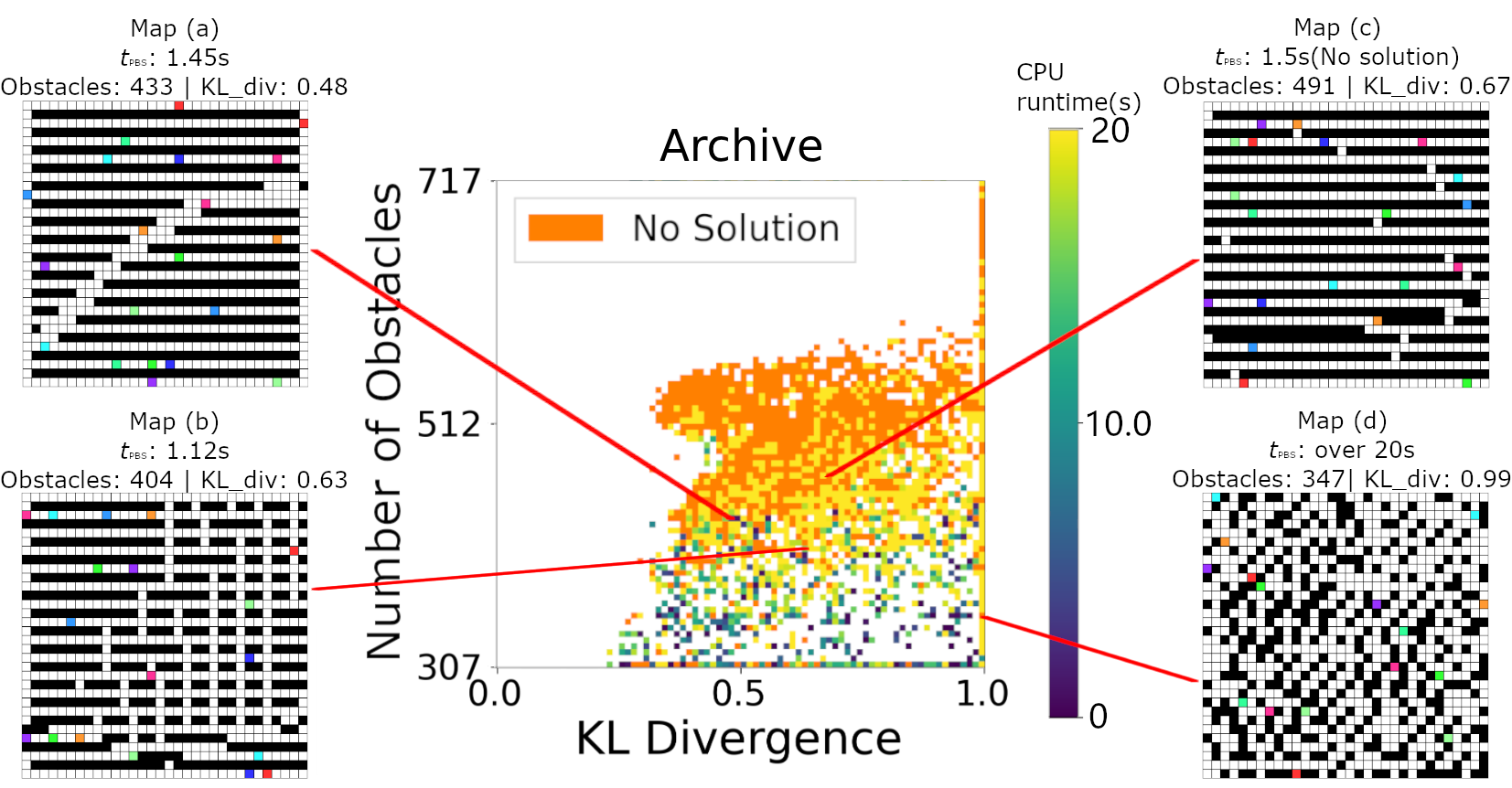}
    \caption{Archive for PBS with sample maps.}
    \Description[Archive of PBS]{Archive of PBS shows the distribution of difficult maps QD-MAPPER generates with four representative maps.}
    \label{fig:PBS}
\end{figure}

\begin{figure}[!t]
    \centering
    \includegraphics[width=0.5\textwidth]{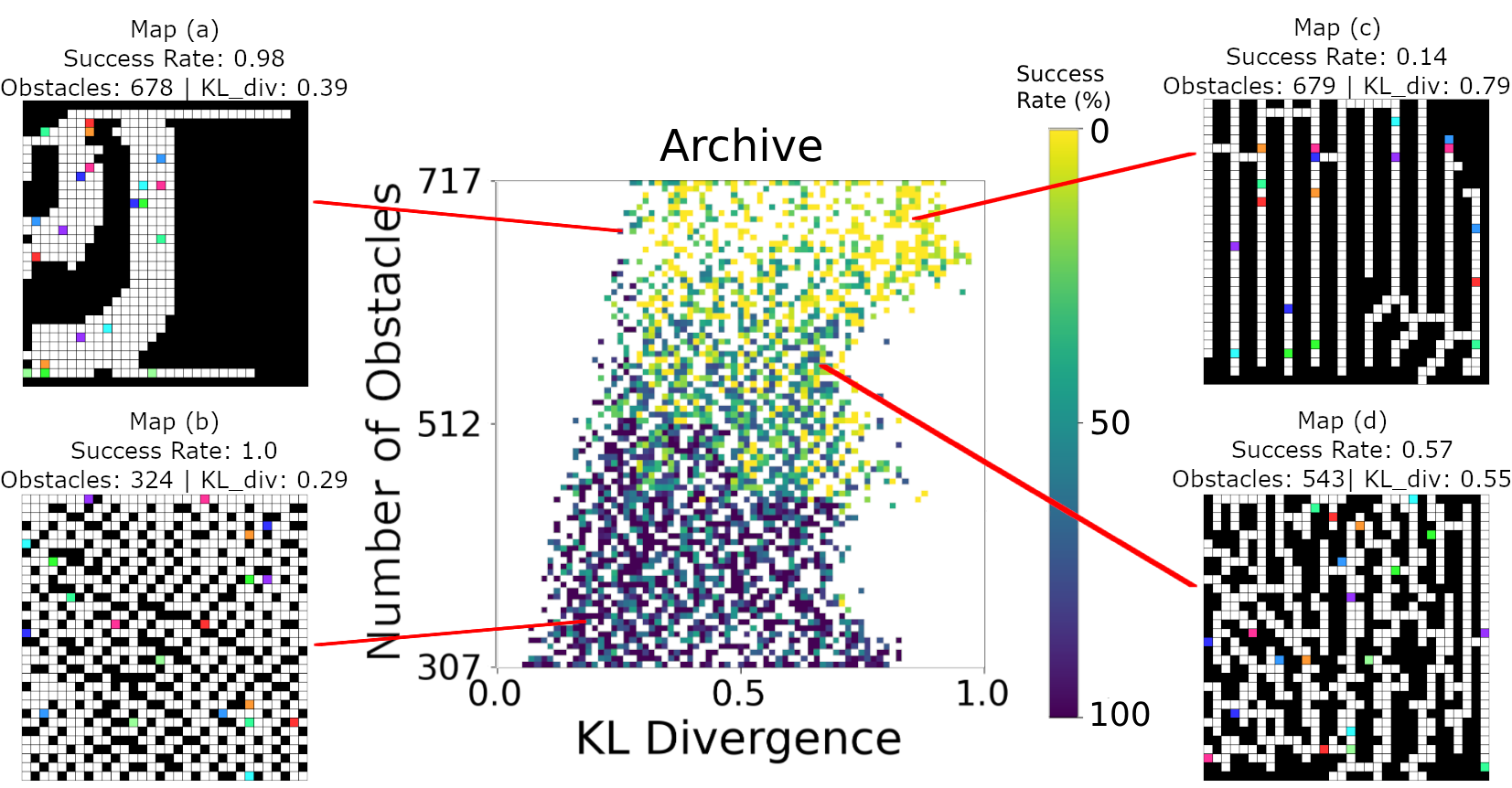}
    \caption{Archive for PIBT with sample maps.}
    \Description[Archive of PIBT]{Archive of PIBT shows the distribution of difficult maps QD-MAPPER generates with four representative maps.}
    \label{fig:PIBT}
\end{figure}

\begin{figure}[!t]
    \centering
    \includegraphics[width=0.5\textwidth]{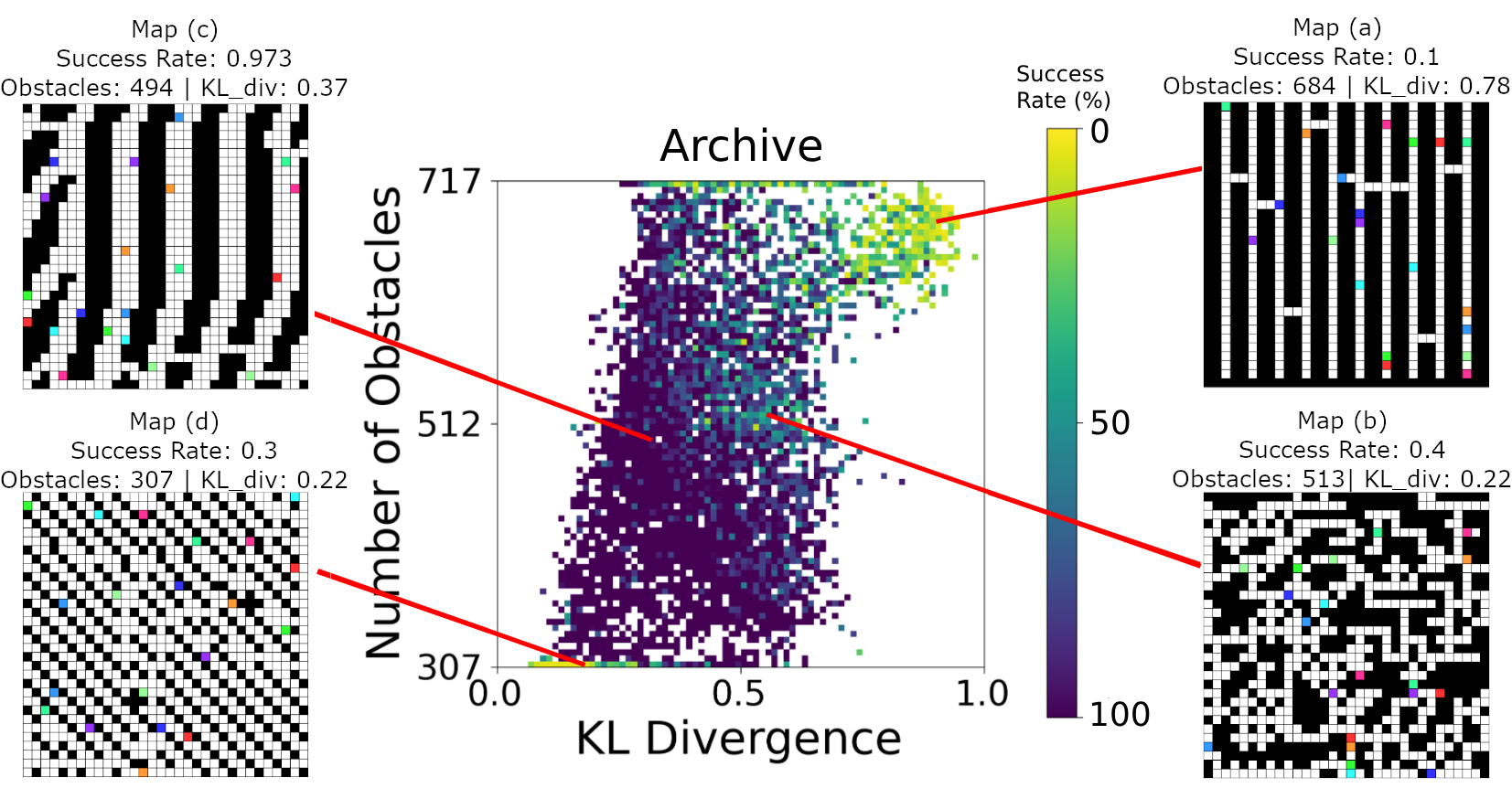}
    \caption{Archive for LTF with sample maps.} 
    \Description[Archive of LTF]{Archive of LTF shows the distribution of difficult maps QD-MAPPER generates with four representative maps.}
    \label{fig:LTF}
\end{figure}

\mysubsubsection{Generated Maps.}
We first define \emph{one-entry space} and \emph{one-tile entry} (shown in \Cref{fig:example-bad}) as features of the maps. A one-entry space represents the empty space surrounded by three obstacles. A one-tile entry represents a single empty space separating two corridors.
In \Cref{fig:CBS,fig:EECBS,fig:PBS}, the yellow and dark blue cells in the archives indicate maps with high and low CPU runtime, respectively.
In \Cref{fig:PIBT,fig:LTF}, the yellow and dark blue cells indicate maps  with low and high success rates, respectively.
The colored tiles in the maps display 10 out of 150 pairs of start and goal locations.
While we are primarily interested in maps that are challenging for the algorithms, we also present easy maps as the by-products of the experiments. We summarize the typical patterns of the maps and show more maps with similar patterns in \Cref{appen:similar_maps}.

\mysubsubsection{CBS.}
\Cref{fig:CBS} shows the archive of maps for CBS. It is easy to generate challenging maps for CBS. Map (a), Map (c), and Map (d) show the most typical hard maps for CBS. Map (a) and Map (c) contain many long corridors, and Map (d) contains many one-entry spaces with long corridors.

Some maps that CBS can solve are maps with fewer obstacles and a large chunk of empty spaces. Map (b) is one of these typical maps. 
Map (b) contains a large chunk of empty spaces and short corridor components in between, providing more space for agents to avoid collisions.


\mysubsubsection{EECBS.}
\Cref{fig:EECBS} shows the archive of maps for EECBS. Compared to the archive of CBS,  it is harder to generate challenging maps for EECBS. 
Map (b) and Map (c) are hard maps with typical patterns for EECBS. 
Map (b) contains a large chunk of empty space, but with many long corridors and one-tile entries.
Map (c) contains many long corridors and one-tile entries. EECBS runs out of time on both maps. 


On the other hand, we observe that EECBS performs well in maps with more empty spaces between each long obstacle component and maps with short obstacle components, where obstacle components refer to clusters of two or more obstacles. Map (a) and Map (d) are two maps with typical patterns. 
Map (a) contains many long obstacle components with one-tile entries; however, the map has two or more columns of empty space between each long obstacle component. 
Map (d) contains many short obstacle components with many entries in between, providing more space for agents to avoid collision. 
Other maps where EECBS performs well are of similar patterns or have a large chunk of empty spaces. 
We show more maps with similar patterns in \Cref{appen:similar_maps}.


\mysubsubsection{PBS.}
\Cref{fig:PBS} shows the archive of maps for PBS.
We empirically discover that PBS returns no solutions in many maps. Therefore, we show the maps in which PBS returns no solution in at least one out of five MAPF instances during the evaluation in the archive as orange cells.
Map (c) and Map (d) are hard maps with typical patterns for PBS. 
Map (c) contains many long corridors with one-tile entries in between, and Map (d) contains many one-entry spaces. PBS always reports no solution on Map (c) and runs out of time on Map (d). 
On the other hand, Map (a) and (b) show two typical patterns where PBS performs well. Map (a) contains long corridors but with more entry spaces in between. Map (b) contains many short obstacle components, providing more space for agents to avoid collisions. 
To validate our observation, we run PBS in 200 MAPF instances on Map (a). The results in \Cref{tab:200-validation} of \Cref{appen:validation} indicate that PBS can effectively solve maps with long corridors with increased entry spaces in between.




\begin{figure}[!t]
    \centering
    \includegraphics[width=0.5\textwidth]{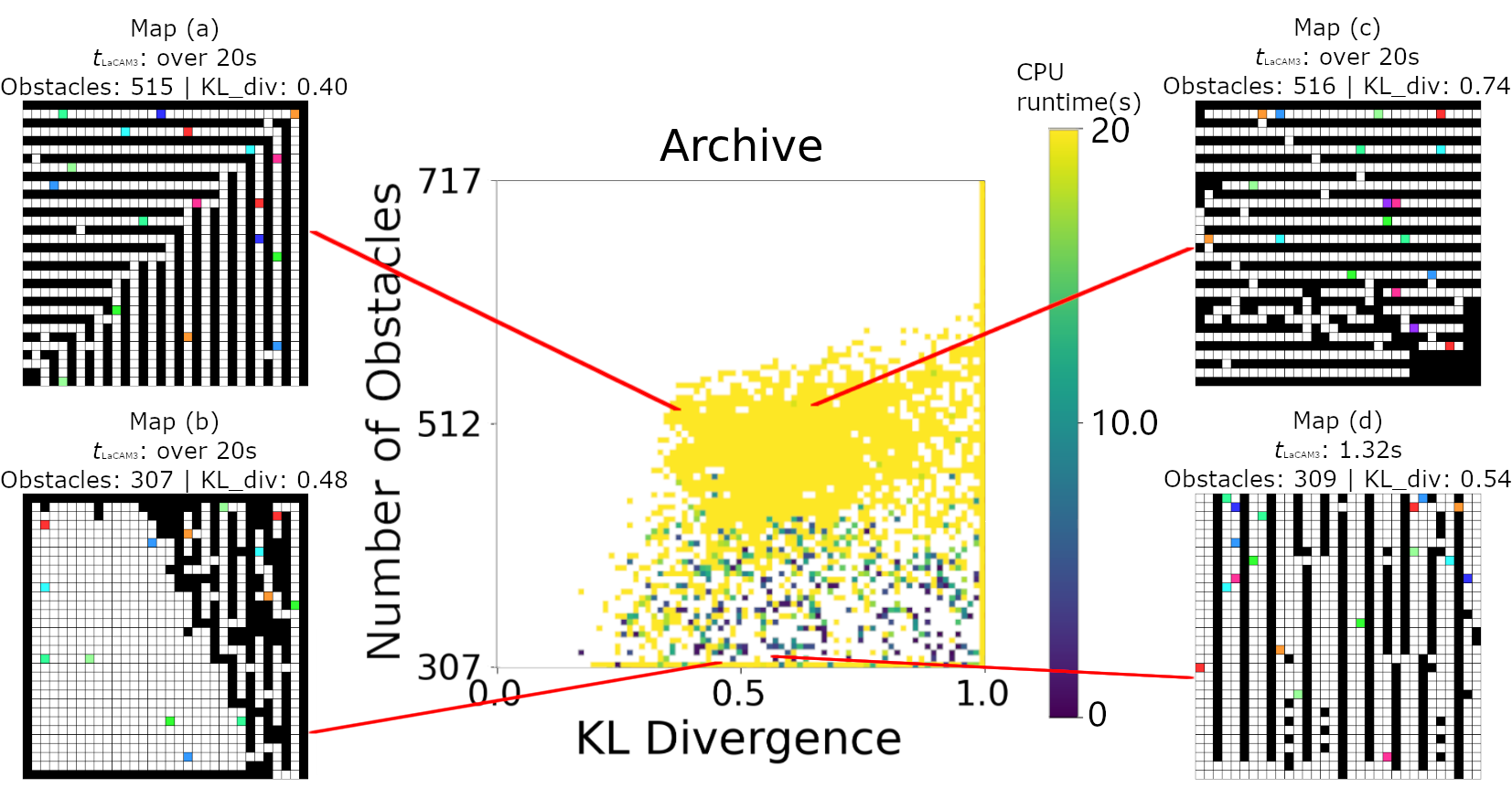}
    \caption{Archive for LaCAM3 with sample maps.}
    \Description[Archive of LaCAM3]{Archive of LaCAM3 shows the distribution of difficult maps QD-MAPPER generates with four representative maps.}
    \label{fig:LaCAM3}
\end{figure}

\mysubsubsection{LaCAM3.}
\Cref{fig:LaCAM3} shows the archive of maps for LaCAM3. 
Notably, the previous work~\cite{okumura2024lacam3} claims that LaCAM3 solves 99\% of the maps in the MAPF benchmark~\cite{SternSoCS19}, while we find a number of challenging maps for the algorithm.
Map (a), Map (b), and Map (c) are hard maps with typical patterns for LaCAM3. 
Map (a) and Map (b) contain many long corridors with one entry.
Map (c) contains a large chunk of empty space, but with many long corridors and one-tile entries.
LaCAM3 runs out of time on all three maps. 
On the other hand, we observe that LaCAM3 performs well in maps with more empty spaces between each long obstacle component, such as Map (d).

\mysubsubsection{PIBT.}
\Cref{fig:PIBT} shows the archive of maps of PIBT.
The success rate of PIBT decreases with an increased number of obstacles and KL divergence.
Similar to EECBS and CBS, PIBT performs poorly in maps with long corridors, such as Map (c).
With a similar number of obstacles and smaller KL divergence, we can find easy maps for PIBT. They have large chunks of empty spaces with few corridors. Map (a) is an example. 
In the middle of the archive, we observe maps with success rates of around 50\%. They usually have long corridors and one-entry spaces. Map (d) is an example. In the bottom left corner, we find Map (b), which has many one-entry spaces but fewer obstacles, achieving a success rate of 100\%.
Overall, PIBT can efficiently solve maps that include long corridors and one-entry spaces with sufficient space.

\mysubsubsection{LTF.} 
\Cref{fig:LTF} shows the archive of maps for LTF. Similar to PIBT, LTF performs worse with more obstacles and higher KL divergence, as shown in Map (a).
Similar to challenging maps for CBS, EECBS, and PIBT, Map (a) contains long corridors. 
With fewer obstacles, LTF performs poorly on maps with many one-entry spaces, such as Map (b) and Map (d). These two maps contain different numbers of obstacles, but they have similar patterns of one-entry spaces. 
On the other hand, LTF performs better with fewer obstacles with large chunks of empty spaces, such as Map (c). 

\mysubsubsection{Validation Results.}
We only evaluate each map by running $N_e = 5$ MAPF instances during the QD search. To further validate our observations on the hardness of the maps, we additionally present \emph{validation results} in \Cref{appen:validation} by running 200 MAPF instances on the selected maps.

\begin{figure}[!t]
    \centering
    \includegraphics[width=0.5\textwidth]{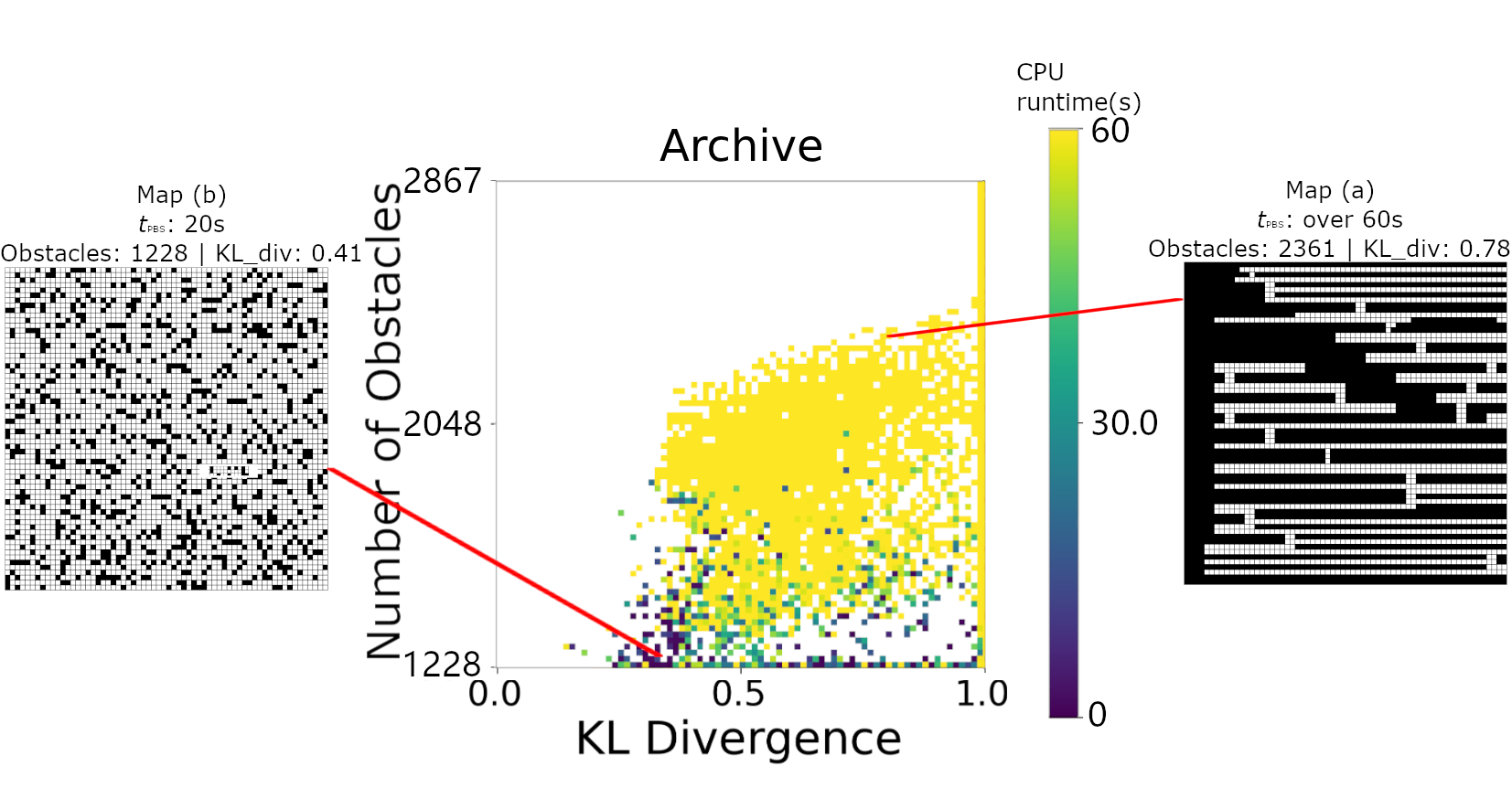}
    \caption{Archive for PBS with sample maps of size 64 $\times$ 64.} 
    \Description[Archive of PBS for size of 64 $\times$ 64]{Archive of PIBT shows the distribution of difficult maps of size 64 $\times$ 64 that QD-MAPPER generates with four representative maps.}
    \label{fig:PBS_64_64}
\end{figure}


\mysubsubsection{On the Generation of Larger Maps.}
To demonstrate that QD-MAPPER can evaluate MAPF algorithms by generating maps larger than 32 $\times$ 32, we further extend our evaluation by generating maps with a size of 64 $\times$ 64 using PBS and PIBT.
\Cref{fig:PBS_64_64} shows the results of PBS. The results of PIBT are shown in \Cref{appen:64}.
Both archives
have a similar distribution compared to experiments of generating benchmark maps with a size of 32 $\times$ 32.

\section{Comparing Two Algorithms} \label{sec:two-algo}

Given the different categories and properties of the MAPF algorithms, hard maps for one algorithm might not be hard for the other. Therefore, we aim to generate maps that are easy for one algorithm and hard for the other by maximizing the performance gap between them.


\begin{figure}[!t]
    \centering
    \includegraphics[width=0.5\textwidth]{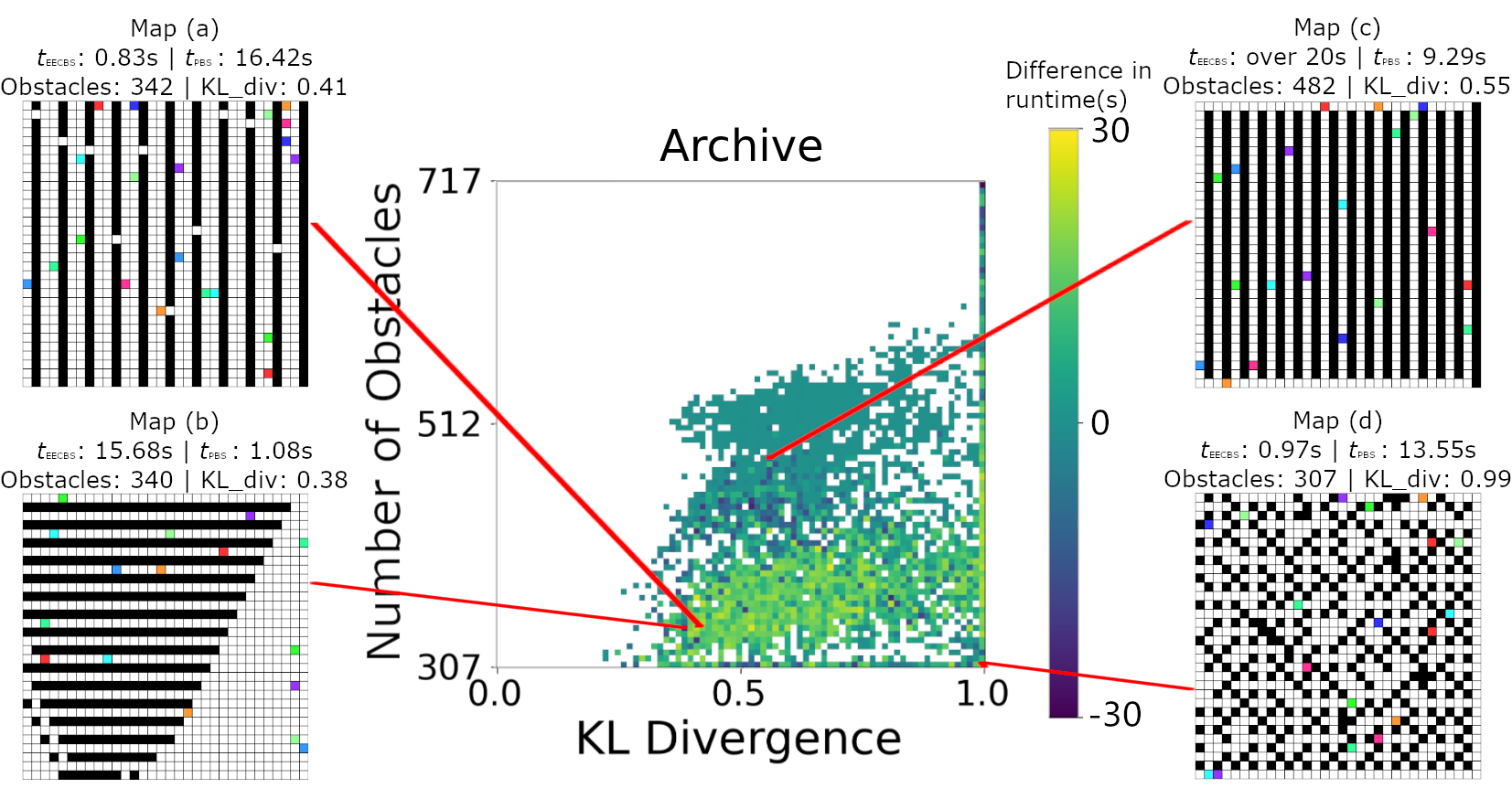}
    \caption{Archive for EECBS vs PBS with sample maps.}
    \Description[Archive of comparison experiment between EECBS and PBS]{Archive of comparison experiment between EECBS and PBS shows the distribution of maps that are easy for one algorithm and hard for the other with four representative maps.}
    \label{fig:eecbs-vs-pbs}
\end{figure}

\begin{figure}[!t]
    \centering
    \includegraphics[width=0.5\textwidth]{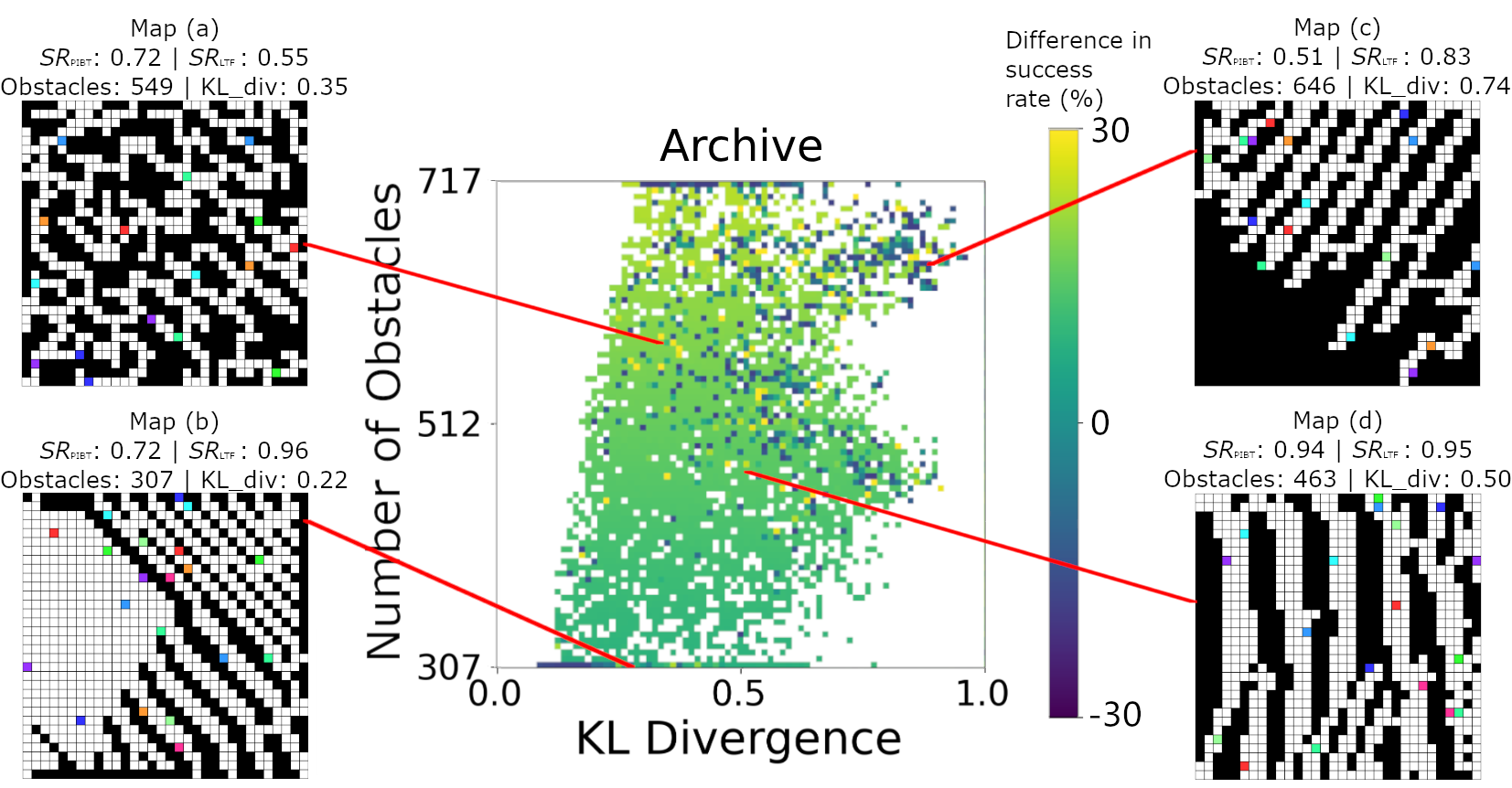}
    \caption{Archive for PIBT vs LTF with sample maps.}
    \Description[Archive of comparison experiment between PIBT and LTF]{Archive of comparison experiment between PIBT and LTF shows the distribution of maps that are easy for one algorithm and hard for the other with four representative maps.}
    \label{fig:pibt-vs-ltf}
\end{figure}

\subsection{Experiment Setup}
We run EECBS with $w = 1.5$. For EECBS and PBS, we use $T = 20$ seconds. For PIBT and LTF, we use $M = 512$, which is used to train the LTF policy~\cite{learntofollow2024}. We use the same $O_{lb}$, $O_{ub}$, $N_{e}$, $N_{eval}$, $N_{a}$, and map sizes as the one-algorithm experiments. 
The compute resources are specified in \Cref{appen:implement-compute}.

\subsection{Results}
\mysubsubsection{Generated Maps.}
We show the archives and representative maps in \Cref{fig:eecbs-vs-pbs,fig:pibt-vs-ltf}. To better visualize the comparisons of the algorithms, we plot the differences, instead of the absolute differences, between the CPU runtime or success rate of the algorithms.
In \Cref{fig:eecbs-vs-pbs}, the yellow and dark blue cells represent EECBS having lower and higher CPU runtime than PBS, respectively.
In \Cref{fig:pibt-vs-ltf}, the yellow and blue cells represent PIBT having higher and lower success rates than LTF, respectively.

\mysubsubsection{EECBS vs. PBS.}
\Cref{fig:eecbs-vs-pbs} shows the archive comparing EECBS and PBS. In most cases, EECBS and PBS have similar performances. 
Maps with more than 512 obstacles are usually too hard for both algorithms. 
With fewer than 400 obstacles, however, EECBS generally performs better than PBS in two types of maps, exemplified in Map (a) and (d). Map (a) is similar to Map (a) of \Cref{fig:EECBS}, with long but wide corridors and one-entry spaces between adjacent corridors.
Map (d) shows a random pattern with many one-entry spaces. 
Meanwhile, Map (b) and (c) are examples where PBS outperforms EECBS. Notably, while Map (c) contains narrow long corridors, both ends of the corridors are open, potentially making PBS resolve conflicts easier. To validate our observation, we run 200 instances on maps shown in \Cref{fig:eecbs-vs-pbs} and show the result in \Cref{tab:200_validation_comp}.
Our results serve as a more comprehensive comparison between EECBS and PBS. Notably, only one work~\cite{Chan2023GreedyPS} has systematically compared EECBS and PBS. They conclude that EECBS outperforms PBS in five out of six maps selected from the MAPF benchmark~\cite{SternSoCS19}. In contrast, our results indicate that EECBS and PBS exhibit similar behavior on most maps, with each having distinct advantages on different map patterns.

\begin{table}[!t]
\centering
\small
\setlength{\tabcolsep}{4pt}
\renewcommand{\arraystretch}{1.1}
\caption{
Validation of observation by running maps with 200 different instances.
\textbf{Left}: average CPU runtime $T$ and success rate $SR$ for EECBS and PBS.
\textbf{Right}: success rate $SR$ for PIBT and LTF.
}
\resizebox{1\linewidth}{!}{
\begin{tabular}{c|rrrr||c|rr}
\toprule
\multicolumn{5}{c||}{\textbf{EECBS vs. PBS}} &
\multicolumn{3}{c}{\textbf{PIBT vs. LTF}} \\
\midrule
Map & $T_{EECBS}$ & $SR_{EECBS}$ & $T_{PBS}$ & $SR_{PBS}$ &
Map & $SR_{PIBT}$ & $SR_{LTF}$ \\
\midrule
Map (a) & 4.3s  & $95.5\%$ & 12.5s & $69.5\%$ & Map (a) & $70\%$ & $56\%$ \\
Map (b) & 9.8s  & $56.5\%$ & 2.0s  & $96.5\%$ & Map (b) & $80\%$ & $96\%$ \\
Map (c) & 18.3s & $14.5\%$ & 13.7s & $43.0\%$ & Map (c) & $52\%$ & $80\%$ \\
Map (d) & 1.1s  & $100.0\%$  & 7.4s  & $77.5\%$ & Map (d) & $96\%$ & $96\%$ \\
\bottomrule
\end{tabular}
}
\label{tab:200_validation_comp}
\end{table}

\mysubsubsection{PIBT vs. LTF.}
\Cref{fig:pibt-vs-ltf} shows the archive that compares PIBT and LTF. 
In most cases with less than 50\% of obstacles, PIBT and LTF have a similar success rate. In maps with large chunks of empty space such as Map (d), both PIBT and LTF perform well. With more than 50\% of obstacles, PIBT generally has a 5\% to 10\% higher success rate on average than LTF. In most cases, these maps contain many one-entry spaces, such as Map (a). 
Upon running 200 instances on Map (a) with PIBT and LTF, PIBT outperforms LTF by an average success rate of 14\%, as shown in \Cref{tab:200_validation_comp}.
The result aligns with our observations in \Cref{sec:one-algo}. 
Meanwhile, the maps in which LTF outperforms PIBT are concentrated in the upper right and bottom left, such as Map (b) and (c).
Upon running 200 instances on both maps, LTF has an average of 20\% higher success rate on these maps compared to PIBT. 
Our results provide additional comparisons between PIBT and LTF to the experiments conducted by \citet{learntofollow2024}, which conclude that LTF outperforms PIBT by comparing them on a set of human-designed warehouse maps.


\section{Discussion and Conclusion} \label{sec:discuss}

We propose QD-MAPPER based on QD algorithms to systematically evaluate the performance of MAPF algorithms by generating diverse maps. We provide two concrete realizations of QD-MAPPER, presenting experimental results with diverse maps of different levels of hardness and patterns. For the one-algorithm experiments, we explore the limitations of five representative MAPF algorithms by generating hard maps by either maximizing the CPU runtime or minimizing the regularized success rate. For the two-algorithm experiments, we compare two algorithms by generating diverse maps that are hard for one algorithm and easy for the other.


\mysubsubsection{Main Observations.}
In one-algorithm experiments, PBS frequently fails to return any solution, not just timing out, on maps with long corridors and sparse or one-entry points, exposing a form of incompleteness that standard benchmark maps have failed to reveal.
LaCAM3, despite prior claims of solving 99\% of standard benchmarks~\cite{okumura2024lacam3}, also struggles under our generated maps with long corridors and narrow entries, which are patterns underrepresented in existing benchmarks. From PIBT–LTF comparison,
as obstacle density rises above 50\%, PIBT typically outperforms LTF by 5–10\%, particularly on maps rich in one-entry spaces. Meanwhile, LTF holds a performance advantage on maps with over 60\% obstacles and high complexity. These findings extend previous evaluations~\cite{learntofollow2024} with LTF and PIBT on a set of human-designed warehouse maps.

\mysubsubsection{When and How to Use QD-MAPPER to Evaluate Performance.}
While developing new MAPF algorithms, researchers can use our one-algorithm pipeline to evaluate their performance by finding maps their algorithm has difficulty solving. In this case, it helps reveal the weakness of new algorithms.
Researchers can also leverage our two-algorithm pipeline to compare their proposed algorithm with an existing algorithm by identifying map types where their algorithm excels or struggles. We argue that this is a more systematic way of comparing two algorithms than comparing on a fixed set of benchmark maps because researchers might intentionally or unintentionally cherry-pick maps that favor their algorithm or overfit the design of their algorithms to the selected set of maps.

QD-MAPPER is a modular framework that separates instance generation, algorithm evaluation, and quality-diversity optimization. 
It allows users to customize in multiple parts.
Researchers can tailor QD-MAPPER for alternative purposes by designing different objectives and diversity measures. 
For example, researchers can 
\emph{minimize} rather than maximize CPU runtime of CBS, EECBS, and PBS with a larger number of agents to generate maps that are \emph{easy}, instead of hard, for these algorithms.
If researchers are interested in both hard and easy maps, using the CPU runtime or regularized success rate as diversity measures is a better choice.
Researchers can also generalize the two-algorithm pipeline to an arbitrary number of MAPF algorithms to rank the performance of more than two algorithms. In addition, researchers can leverage learning-based diversity measures~\cite{Li2024_DiversityHumanFeedback,wang2025diversityhumanfeedback} to evaluate MAPF algorithms on maps that meet criteria defined by humans. 
Future research can also extend QD‑MAPPER by generating \emph{instances} with varied numbers of agents, map sizes, as well as start and goal locations to provide a comprehensive evaluation of algorithm behavior under diverse configuration settings.





\section*{Acknowledgments}
This work used Bridge-$2$ at Pittsburgh Supercomputing Center (PSC) through allocation CIS$220115$ from the Advanced Cyberinfrastructure Coordination Ecosystem: Services \& Support (ACCESS) program, which is supported by National Science Foundation grants \#$2138259$, \#$2138286$, \#$2138307$, \#$2137603$, and \#$2138296$.

\bibliographystyle{ACM-Reference-Format} 
\bibliography{sample}

@ARTICLE{guillaume2018primal,
  author={Sartoretti, Guillaume and Kerr, Justin and Shi, Yunfei and Wagner, Glenn and Kumar, T. K. Satish and Koenig, Sven and Choset, Howie},
  journal={IEEE Robotics and Automation Letters}, 
  title={PRIMAL: Pathfinding via Reinforcement and Imitation Multi-Agent Learning}, 
  year={2019},
  volume={4},
  number={3},
  pages={2378-2385},
}

@Inproceedings{Wang2023SCRIMPSC,
  title={SCRIMP: Scalable Communication for Reinforcement- and Imitation-Learning-Based Multi-Agent Pathfinding},
  author={Yutong Wang and Bairan Xiang and Shinan Huang and Guillaume Sartoretti},
  booktitle={IROS},
  year={2023},
  pages={9301-9308},
}

@Inproceedings{Andreychuk_MAPFGPT_2025, title={MAPF-GPT: Imitation Learning for Multi-Agent Pathfinding at Scale}, volume="", url="", DOI="", number={22}, booktitle={AAAI}, author={Andreychuk, Anton and Yakovlev, Konstantin and Panov, Aleksandr and Skrynnik, Alexey}, year={2025}, month="", pages={23126-23134} }

@inproceedings{ JiangICRA25,
  author    = "He Jiang and Yutong Wang and Rishi Veerapaneni and Tanishq Harish Duhan and Guillaume Adrien Sartoretti and Jiaoyang Li",
  title     = "Deploying Ten Thousand Robots: Scalable Imitation Learning for Lifelong Multi-Agent Path Finding",
  booktitle = "ICRA",
  pages     = "1--7",
  year      = "2025",
  doi       = "",
}

@inproceedings{
skrynnik2025pogema,
title={{POGEMA}: A Benchmark Platform for Cooperative Multi-Agent Pathfinding},
author={Alexey Skrynnik and Anton Andreychuk and Anatolii Borzilov and Alexander Chernyavskiy and Konstantin Yakovlev and Aleksandr Panov},
booktitle={ICLR},
year={2025},
url=""
}

@inproceedings{Jiang2024Competition,
    author    = {He Jiang and Yulun Zhang and Rishi Veerapaneni and Jiaoyang Li},
    title     = {Scaling Lifelong Multi-Agent Path Finding to More Realistic Settings: Research Challenges and Opportunities},
    booktitle = {SoCS},
    pages     = {234--242},
    year      = {2024},
    doi       = ""
}

@inproceedings{Li2024_DiversityHumanFeedback,
author = {Ding, Li and Zhang, Jenny and Clune, Jeff and Spector, Lee and Lehman, Joel},
title = {Quality Diversity Through Human Feedback: Towards Open-Ended Diversity-Driven Optimization},
year = {2024},
publisher = "",
booktitle = {ICML},
pages = {11072--11090},
numpages = "",
location = "",
}

@article{wang2025diversityhumanfeedback,
author={Wang, Ren-Jian
and Xue, Ke
and Wang, Yu-Tong
and Yang, Peng
and Fu, Hao-Bo
and Fu, Qiang
and Qian, Chao},
title={Diversity from Human Feedback},
journal={Frontiers of Computer Science},
year={2025},
volume={20},
number={2},
pages={2002320},
issn={2095-2236},
}

@inproceedings{StandleyAAAI10,
  author    = {Trevor Scott Standley},
  title     = {Finding Optimal Solutions to Cooperative Pathfinding Problems},
  booktitle = {AAAI},
    pages={173--178},
  year      = {2010}
}

@article{PearTPAMIl1982,
  title={Studies in Semi-Admissible Heuristics},
  author={Pearl, Judea and Kim, Jin H},
  journal={IEEE Transactions on Pattern Analysis and Machine Intelligence},
  volume={4}, 
  number={4},
  pages={392--399},
  year={1982},
  publisher={IEEE}
}

@inproceedings{LamIJCAI19,
  author    = {Edward Lam and
               Le Bodic, Pierre and
               Daniel Harabor and
               Peter J. Stuckey},
  title     = {Branch-and-Cut-and-Price for Multi-Agent Pathfinding},
  booktitle = {IJCAI},
  pages     = {1289--1296},
  year      = {2019}
}

@inproceedings{MaAAAI19,
  author    = {Hang Ma and
               Daniel Harabor and
               Peter J. Stuckey and
               Jiaoyang Li and
               Sven Koenig},
  title     = {Searching with Consistent Prioritization for Multi-Agent Path Finding},
  booktitle = {AAAI},
  pages     = {7643--7650},
  year      = {2019}
}

@inproceedings{SternSoCS19,
  author    = {Roni Stern and
               Nathan R. Sturtevant and
               Ariel Felner and
               Sven Koenig and
               Hang Ma and
               Thayne T. Walker and
               Jiaoyang Li and
               Dor Atzmon and
               Liron Cohen and
               T. K. Satish Kumar and
               Roman Bart{\'{a}}k and
               Eli Boyarski},
  title     = {Multi-Agent Pathfinding: Definitions, Variants, and Benchmarks},
  booktitle = {SoCS},
  pages     = {151--159},
  year      = {2019}
}

@article{SharonAIJ15,
  author    = {Guni Sharon and
               Roni Stern and
               Ariel Felner and
               Nathan R. Sturtevant},
  title     = {Conflict-Based Search for Optimal Multi-Agent Pathfinding},
  journal   = {Artificial Intelligence},
  volume    = {219},
  pages     = {40--66},
  year      = {2015}
}

@inproceedings{pyribs,
  author    = {Bryon Tjanaka and Matthew C. Fontaine and David H. Lee and Yulun Zhang and Nivedit Reddy Balam and Nathaniel Dennler and Sujay S. Garlanka and Nikitas Dimitri Klapsis and Stefanos Nikolaidis },
  title     = {pyribs: A Bare-Bones Python Library for Quality Diversity Optimization},
  booktitle = {GECCO},
  pages     = {220--229},
  year      = {2023}
}

@inproceedings{Li2020EECBSAB,
  title={EECBS: A Bounded-Suboptimal Search for Multi-Agent Path Finding},
  author={Jiaoyang Li and Wheeler Ruml and Sven Koenig},
  booktitle={AAAI},
  year={2021},
  pages={12353--12362}
}

@article{SharonAIJ13,
  author    = {Guni Sharon and
               Roni Stern and
               Meir Goldenberg and
               Ariel Felner},
  title     = {The Increasing Cost Tree Search for Optimal Multi-Agent Pathfinding},
  journal   = {Artificial Intelligence},
  volume    = {195},
  pages     = {470--495},
  year      = {2013}
}

@article{sturtevant2012benchmarks,
  title={Benchmarks for Grid-Based Pathfinding},
  author={Sturtevant, N.},
  journal={Transactions on Computational Intelligence and AI in Games},
  volume={4},
  number={2},
  pages={144--148},
  year={2012},
}

@article{MStar,
  author    = {G. Wagner and
               H. Choset},
  title     = {Subdimensional Expansion for Multirobot Path Planning},
  journal   = {Artificial Intelligence},
  volume    = {219},
  pages     = {1--24},
  year      = {2015}
}

@inproceedings{BarrerSoCS14,
  author    = {Max Barer and
               Guni Sharon and
               Roni Stern and
               Ariel Felner},
  title     = {Suboptimal Variants of the Conflict-Based Search Algorithm for the Multi-Agent Pathfinding Problem},
  booktitle = {SoCS},
  pages = "19--27",
  year      = {2014}
}

@article{Erdmann87,
  author    = {Michael A. Erdmann and
               Tom{\'{a}}s Lozano{-}P{\'{e}}rez},
  title     = {On Multiple Moving Objects},
  journal   = {Algorithmica},
  volume    = {2},
  pages     = {477--521},
  year      = {1987}
}

@article{WangB11,
  author    = {K. Wang and
               A. Botea},
  title     = {{MAPP}: A Scalable Multi-Agent Path Planning Algorithm with Tractability
               and Completeness Guarantees},
  journal   = {Journal of Artificial Intelligence Research},
  volume    = {42},
  pages     = {55--90},
  year      = {2011}
}

@inproceedings{MaAIIDE17,
	author    = {Hang Ma and
               Jingxing Yang and
               Liron Cohen and
               T. K. Satish Kumar and
               Sven Koenig},
  title     = {Feasibility Study: Moving Non-Homogeneous Teams in Congested Video Game Environments},
  booktitle = {AIIDE},
    pages = {270--272},
	year = 2017
}

@inproceedings{fontaine2021quality,
    title={A Quality Diversity Approach to Automatically Generating Human-Robot
           Interaction Scenarios in Shared Autonomy},
    author={Matthew C. Fontaine and Stefanos Nikolaidis},
    year={2021},
    url="",
    booktitle={RSS},
    doi="",
}

@inproceedings{
fontaine2020illuminating, 
title={Illuminating {Mario} Scenes in the Latent Space of a Generative Adversarial Network},
url="",
author={Fontaine, Matthew C. and Liu, Ruilin and Khalifa, Ahmed and Modi, Jignesh and Togelius, Julian and Hoover, Amy K. and Nikolaidis, Stefanos}, 
booktitle = {AAAI},
year={2021}, 
pages={5922--5930} }

@inproceedings{Fontaine2022CovarianceMA,
author = {Fontaine, Matthew and Nikolaidis, Stefanos},
title = {Covariance Matrix Adaptation MAP-Annealing},
year = {2023},
booktitle = {GECCO},
pages = {456--465},
}

@inproceedings{zhang:aiide2020,
author = {Zhang, Hejia and Fontaine, Matthew C. and Hoover, Amy K. and Togelius, Julian and Dilkina, Bistra and Nikolaidis, Stefanos},
title = {Video Game Level Repair via Mixed Integer Linear Programming},
year = {2020},
booktitle = {AIIDE},
pages={151--158}
}

@article{hansen2016cmaes,
  title={The {CMA} Evolution Strategy: A Tutorial},
  author={Nikolaus Hansen},
  journal={ArXiv},
  year={2016},
  volume={abs/1604.00772},
  url=""
}

@article{Wang2019PairedOT,
  title={Paired Open-Ended Trailblazer (POET): Endlessly Generating Increasingly Complex and Diverse Learning Environments and Their Solutions},
  author={Rui Wang and Joel Lehman and Jeff Clune and Kenneth O. Stanley},
  journal={ArXiv},
  year={2019},
  volume={abs/1901.01753},
  pages={},
}

@inproceedings{Cobbe2019LeveragingPG,
author = {Cobbe, Karl and Hesse, Christopher and Hilton, Jacob and Schulman, John},
title = {Leveraging procedural generation to benchmark reinforcement learning},
year = {2020},
booktitle = {ICML},
pages={2048--2056}
}

@inproceedings{Abeysirigoonawardena2019GeneratingAD,
  title={Generating Adversarial Driving Scenarios in High-Fidelity Simulators},
  author={Yasasa Abeysirigoonawardena and Florian Shkurti and Gregory Dudek},
  booktitle={ICRA},
  year={2019},
  pages={8271--8277}
}

@article{Mullins2018AdaptiveGO,
  title={Adaptive Generation of Challenging Scenarios for Testing and Evaluation of Autonomous Vehicles},
  author={Galen E. Mullins and Paul G. Stankiewicz and R. Chad Hawthorne and Satyandra K. Gupta},
  journal={Journal of Systems and Software},
  year={2018},
  volume={137},
  pages={197--215}
}

@article{gardner1970,
  title={Mathematical Games – The Fantastic Combinations of John Conway's New Solitaire Game ``Life''},
  author={Gardner, Martin},
  journal={Scientific American},
  volume={223},
  number={4},
  pages={120--123},
  year={1970}
}

@inproceedings{fontaine2021importance,
    title={On the Importance of Environments in Human-Robot Coordination},
    author={Matthew C. Fontaine and Ya-Chuan Hsu and Yulun Zhang and Bryon Tjanaka and Stefanos Nikolaidis},
    year={2021},
    url="",
    booktitle={RSS},
    doi="",
    pages = {}
}

@inproceedings{Zhang2021DeepSA,
author = {Zhang, Yulun and Fontaine, Matthew C. and Hoover, Amy K. and Nikolaidis, Stefanos},
title = {Deep Surrogate Assisted MAP-Elites for Automated Hearthstone Deckbuilding},
year = {2022},
url = "",
doi = "",
booktitle = {GECCO},
pages = {158--167},
numpages = {10},
location = "",
}

@inproceedings{Bhatt2022DeepSA,
  author = {Bhatt, Varun and Tjanaka, Bryon and Fontaine, Matthew and Nikolaidis, Stefanos},
  booktitle = {NeurIPS},
  pages = {37762--37777},
  title = {Deep Surrogate Assisted Generation of Environments},
  year = {2022}
}

@inproceedings{ZhangNCA2023,
  author    = {Yulun Zhang and Matthew C. Fontaine and Varun Bhatt and Stefanos Nikolaidis and Jiaoyang Li},
  title     = {Arbitrarily Scalable Environment Generators via Neural Cellular Automata},
  booktitle = {NeurIPS},
  pages     = {57212--57225},
  year      = {2023}
}

@inproceedings{zhangLayout23,
  author    = {Yulun Zhang and Matthew C. Fontaine and Varun Bhatt and Stefanos Nikolaidis and Jiaoyang Li},
  title     = {Multi-Robot Coordination and Layout Design for Automated Warehousing},
  booktitle = {IJCAI},
  pages     = {5503--5511},
  year      = {2023},
  doi       = ""
}

@article{mouret2015illuminating,
  title={Illuminating Search Spaces by Mapping Elites},
  author={Jean-Baptiste Mouret and Jeff Clune},
  journal={ArXiv},
  year={2015},
  volume={abs/1504.04909}
}

@inproceedings{okumura2019priority,
  title={Priority Inheritance with Backtracking for Iterative Multi-agent Path Finding},
  author={Okumura, Keisuke and Machida, Manao and D{\'e}fago, Xavier and Tamura, Yasumasa},
  booktitle={IJCAI},
  pages={535--542},
  year={2019},
}

@inproceedings{choudhury2022TruckUAVmapf,
author = {Choudhury, Shushman and Solovey, Kiril and Kochenderfer, Mykel and Pavone, Marco},
title = {Coordinated Multi-Agent Pathfinding for Drones and Trucks over Road Networks},
year = {2022},
booktitle = {AAMAS},
pages = {272--280},
}

@inproceedings{ShaoulICAPS24,
  author    = {Yorai Shaoul and Itamar Mishani and Maxim Likhachev and Jiaoyang Li},
  title     = {Accelerating Search-Based Planning for Multi-Robot Manipulation by Leveraging Online-Generated Experiences},
  booktitle = {ICAPS},
  pages     = {523--531},
  doi       = "",
  year      = {2024}
}

@inproceedings{learntofollow2024,
title={Learn to Follow: Decentralized Lifelong Multi-Agent Pathfinding via Planning and Learning},
author={Skrynnik, Alexey and Andreychuk, Anton and Nesterova, Maria and Yakovlev, Konstantin and Panov, Aleksandr},
booktitle = {AAAI},
pages={17541-17549},
year={2024},
}

@inproceedings{friedrich2024scalable,
  author    = {Paul Friedrich and Yulun Zhang and Michael Curry and Ludwig Dierks and Stephen McAleer and Jiaoyang Li and Tuomas Sandholm and Sven Seuken},
  title     = {Scalable Mechanism Design for Multi-Agent Path Finding},
  booktitle = {IJCAI},
  pages     = {58--66},
  year      = {2024},
  doi       = ""
}

@inproceedings{ren2024MAPFhard,
author={Ren, Jingyao and Ewing, Eric and Kumar, T. K. Satish and Koenig, Sven and Ayanian, Nora}, 
title={Map Connectivity and Empirical Hardness of Grid-based Multi-Agent Pathfinding Problem},   
booktitle={ICAPS}, 
pages={484-488}, 
year={2024}, 
}

@inproceedings{Bhatt2023surrHRI,
  author = {Bhatt, Varun and Nemlekar, Heramb and Fontaine, Matthew Christopher and Tjanaka, Bryon and Zhang, Hejia and Hsu, Ya-Chuan and Nikolaidis, Stefanos},
  booktitle = {CoRL},
  pages = {513-539},
  title = {Surrogate Assisted Generation of Human-Robot Interaction Scenarios},
  year = 2023
}

@inproceedings{Chan2023GreedyPS,
  title={Greedy Priority-Based Search for Suboptimal Multi-Agent Path Finding},
  author={Shao-Hung Chan and Roni Stern and Ariel Felner and Sven Koenig},
  booktitle={SoCS},
  year={2023},
  pages = {11--19},

}

@inproceedings{ewing2022BC,
author = {Ewing, Eric and Ren, Jingyao and Kansara, Dhvani and Sathiyanarayanan, Vikraman and Ayanian, Nora},
title = {Betweenness Centrality in Multi-Agent Path Finding},
year = {2022},
booktitle = {AAMAS},
pages = {400–408},
}

@article{Nino2011WLkernal,
author = {Shervashidze, Nino and Schweitzer, Pascal and van Leeuwen, Erik Jan and Mehlhorn, Kurt and Borgwardt, Karsten M.},
title = {Weisfeiler-Lehman Graph Kernels},
year = {2011},
issue_date = {2/1/2011},
volume = {12},
number = {77},
journal = {Journal of Machine Learning Research},
pages = {2539–2561},
numpages = {23}
}

@inproceedings{zhang2024ggo,
  author    = {Yulun Zhang and He Jiang and Varun Bhatt and Stefanos Nikolaidis and Jiaoyang Li},
  title     = {Guidance Graph Optimization for Lifelong Multi-Agent Path Finding},
  booktitle = {IJCAI},
  pages     = {311--320},
  year      = {2024},
}

@inproceedings{Thayer2011EES,
author = {Thayer, Jordan T. and Ruml, Wheeler},
title = {Bounded Suboptimal Search: A Direct Approach Using Inadmissible Estimates},
year = {2011},
booktitle = {IJCAI},
pages = {674–679},
}

@inproceedings{okumura2024lacam3,
  title = {Engineering LaCAM$^\ast$: Towards Real-Time, Large-Scale, and Near-Optimal Multi-Agent Pathfinding},
  booktitle={AAMAS},
  year={2024},
  author={Okumura, Keisuke},
  pages={1501--1509}
}

@inproceedings{PSCBridgeTwo2021,
author = {Brown, Shawn T. and Buitrago, Paola and Hanna, Edward and Sanielevici, Sergiu and Scibek, Robin and Nystrom, Nicholas A.},
title = {Bridges-2: A Platform for Rapidly-Evolving and Data Intensive ResearchBridges-2: A Platform for Rapidly-Evolving and Data Intensive Research},
year = {2021},
booktitle = {PEARC},
pages = {1--4},
}

\clearpage

\appendix


\section{NCA Generation Process} \label{appen:nca}

\begin{figure*}[!t]
    \centering
    \begin{subfigure}{0.19\textwidth}
        \includegraphics[width=1\textwidth]{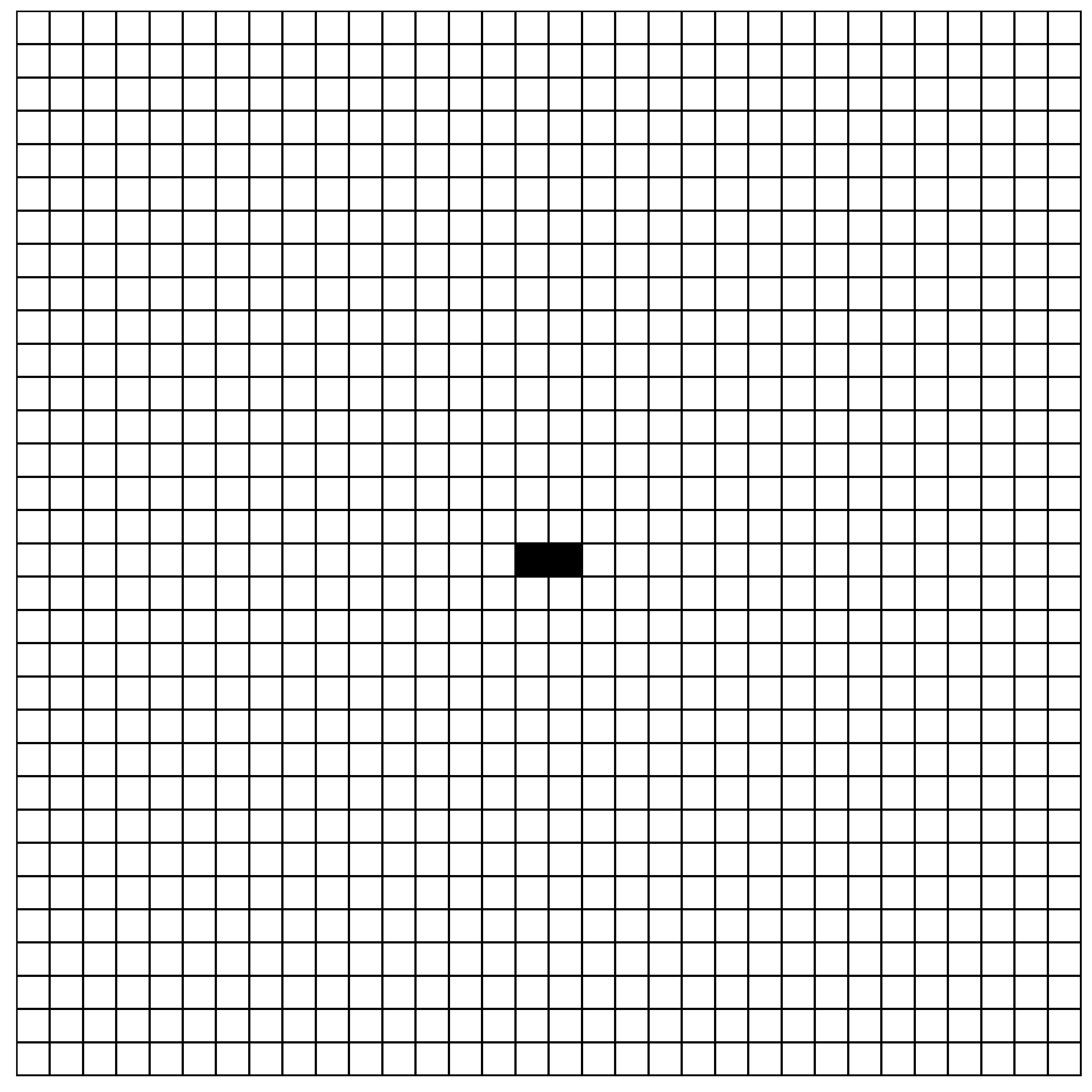}
        \Description{NCA map at iteration 0.}
        \caption{Iteration 0}
        \label{fig:nca-seed}
    \end{subfigure}
    \hfill
    \begin{subfigure}{0.19\textwidth}
        \includegraphics[width=1\textwidth]{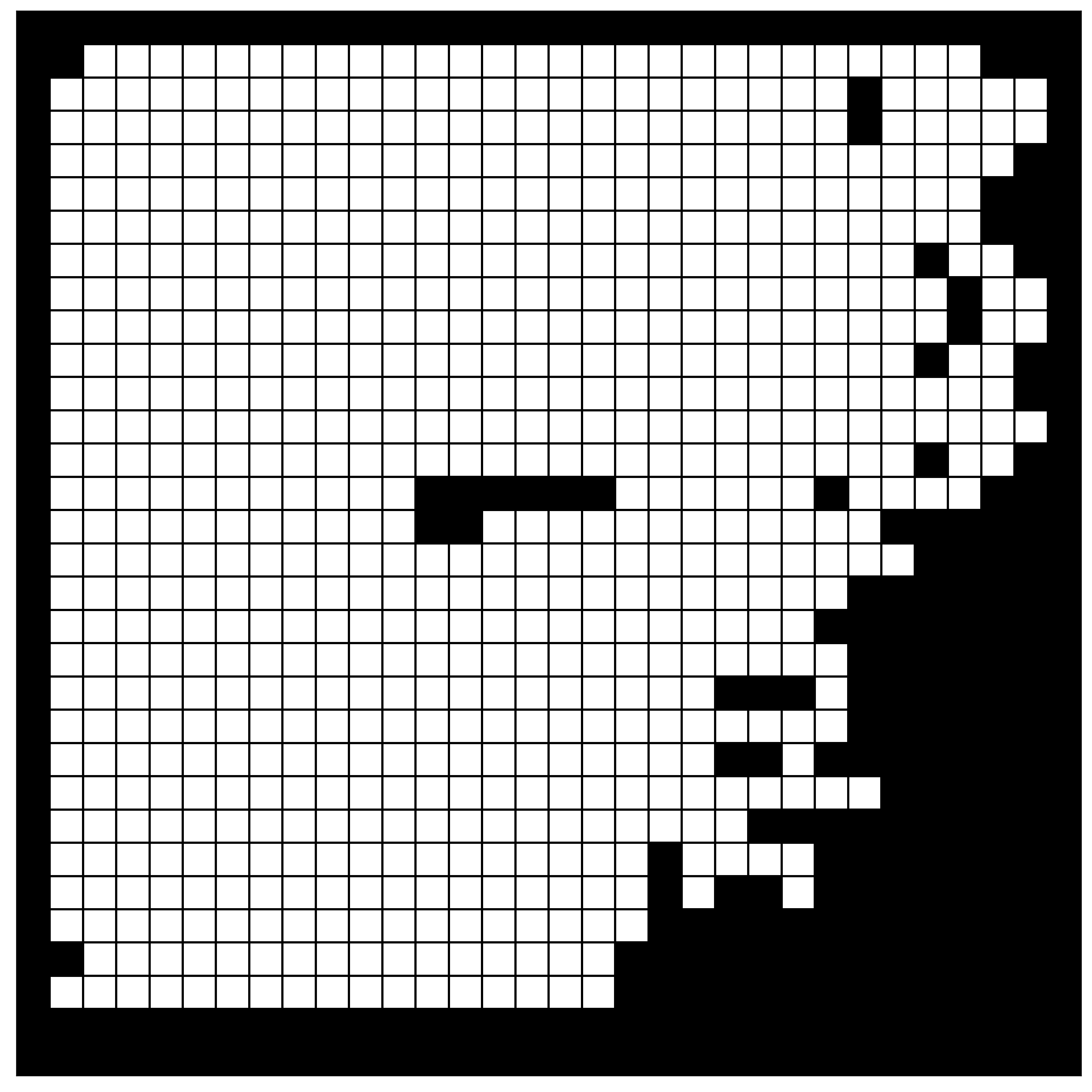}
        \Description{NCA map at iteration 10.}
        \caption{Iteration 10}
        \label{fig:nca-iter5}
    \end{subfigure}
    \hfill
    \begin{subfigure}{0.19\textwidth}
        \includegraphics[width=1\textwidth]{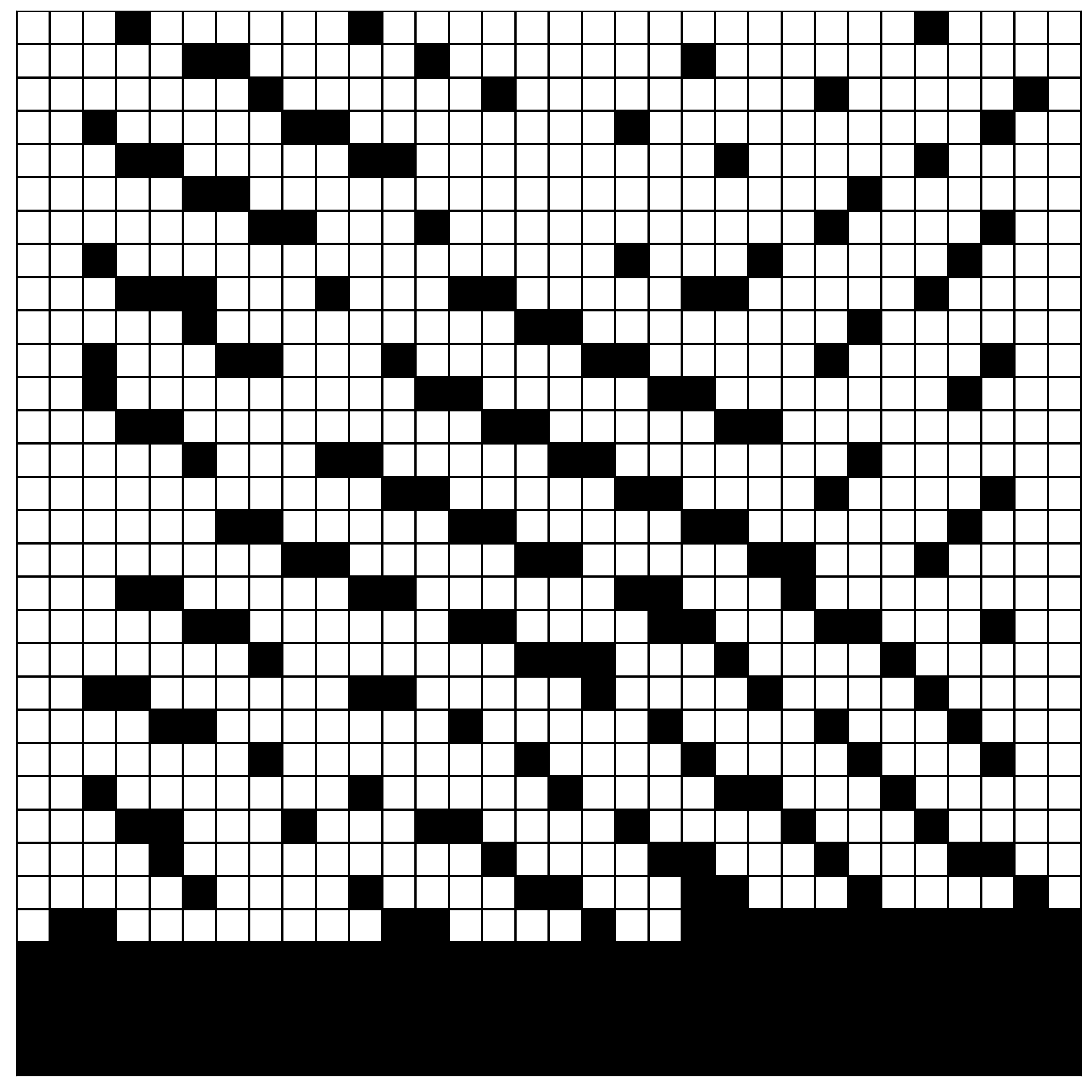}
        \Description{NCA map at iteration 20.}
        \caption{Iteration 20}
        \label{fig:nca-iter10}
    \end{subfigure}
    \hfill
    \begin{subfigure}{0.19\textwidth}
        \includegraphics[width=1\textwidth]{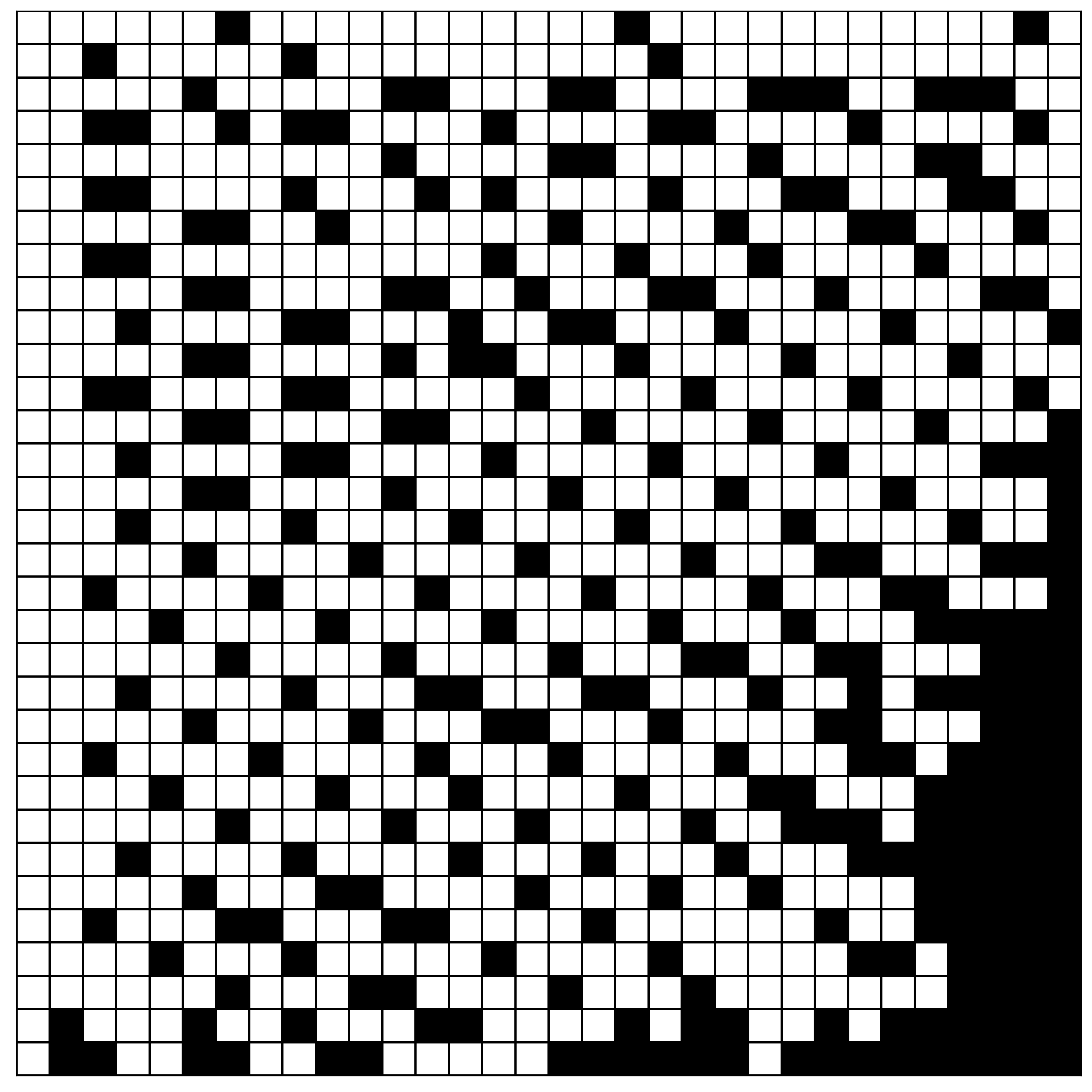}
        \Description{NCA map at iteration 30.}
        \caption{Iteration 30}
        \label{fig:nca-iter30}
    \end{subfigure}
    \hfill
    \begin{subfigure}{0.19\textwidth}
        \includegraphics[width=1\textwidth]{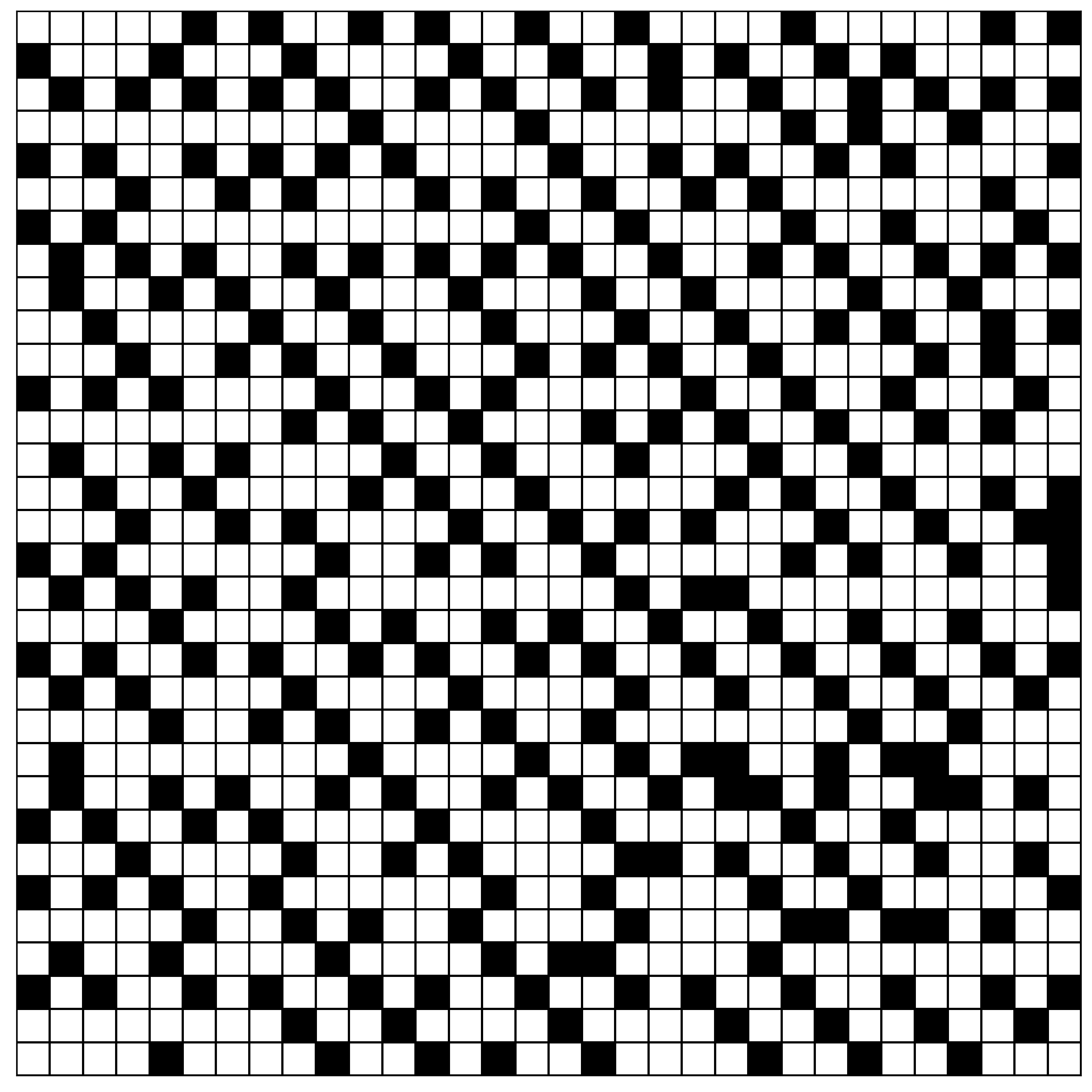}
        \Description{NCA map at iteration 50.}
        \caption{Iteration 50}
        \label{fig:nca-iter50}
    \end{subfigure}
    \caption{
    Example of NCA map generation process with 50 iterations, starting from a fixed simple map (\Cref{fig:nca-seed}) and iteratively generating maps with complex patterns (\Cref{fig:nca-iter5,fig:nca-iter10,fig:nca-iter30,fig:nca-iter50}).}
    \label{fig:nca}
\end{figure*}

\Cref{fig:nca} exemplifies the process for an NCA generator to generate a map in 50 iterations.

\section{Choosing Diversity Measures} \label{appen:measures}

In our experiments, we choose the number of obstacles and KL divergence of tile pattern distribution as the diversity measures. However, we also explore other diversity measures, both in terms of General Hardness and Spatial Arrangement. 
In this section, we first introduce other diversity measures that we have explored. We then conduct experiments to justify our choice of diversity measures.


\begin{figure}[!t]
    \centering
    \includegraphics[width=0.4\textwidth]{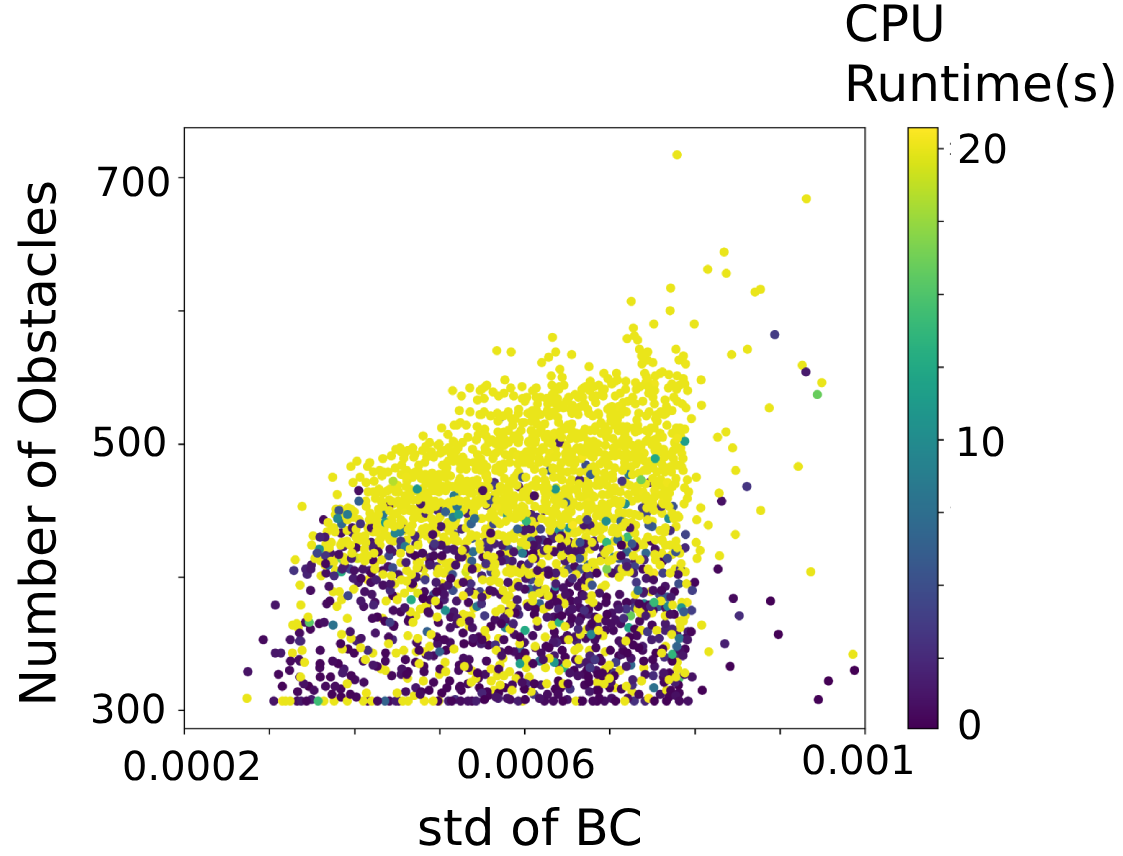}
    \Description{Scatter plot of about 4370 generated maps.}
    \caption{Scatter plot of about 4370 generated maps.}
    \label{fig:obs_bc}
\end{figure}

\subsection{General Hardness} \label{appen:diversity:hard}

\mysubsubsection{Standard Deviation of Betweenness Centrality (Std of BC).} 
Betweenness centrality is claimed to affect the empirical difficulty of MAPF instances~\cite{ewing2022BC}. The Betweenness Centrality of an empty space is the fraction of all possible shortest paths in the map that pass through it. We use the standard deviation of Betweenness Centrality here to measure the congestion in the map.
To compute the std of BC, we search for the shortest path between every pair of empty spaces in the map and then calculate the standard deviation of the usage of all empty spaces. 
The higher the std of BC, the more congestion and the harder the MAPF problem.

\mysubsubsection{$\boldsymbol{\lambda_2}$.}
The second smallest eigenvalue (known as $\lambda_2$) of the normalized Laplacian matrix of different maps is claimed to be correlated to the difficulty of the maps~\cite{ren2024MAPFhard}. 
To compute $\lambda_2$ of a map, we view the map as a graph, and compute the second-smallest eigenvalue of the normalized Laplacian matrix of the graph. 
\citet{ren2024MAPFhard} shows empirically that $\lambda_2$ is correlated with the hardness of the map.

\mysubsubsection{Comparison Experiments: Number of Obstacles vs. Std of BC.}
To compare the ability of quantifying the general hardness of the maps between the number of obstacles and the std of BC, we generate 4370 maps with QD-MAPPER with the objective being the similarity score~\cite{ZhangNCA2023} between repaired and unrepaired maps of the MILP solver. We then set the diversity measure as the number of obstacles and std of BC. In this way, we generate a collection of maps that are not biased towards any algorithm while making sure that they are diverse in terms of the two measures of interest.
Then we run PBS on all the generated maps, collecting average CPU runtime over $N_e = 5$ MAPF instances and the std of BC. We plot them as a scatter plot in \Cref{fig:obs_bc}.
We observe that in the y-axis, with the increase of obstacles, the CPU runtime of PBS is increasing, which is consistent with the one-algorithm experiment result of PBS in \Cref{sec:one-algo} of the main text. With fewer variations in the CPU runtime of PBS at a given number of obstacles, we can quantify the general hardness of the maps based on the number of obstacles. 
From the x-axis, however, there are many maps with the same std of BC but showing different CPU runtimes of the maps for PBS. 
Therefore, we argue that the number of obstacles is a better measure than the std of BC to quantify the general hardness of the maps.

\begin{figure}[!t]
    \centering
    \includegraphics[width=0.4\textwidth]{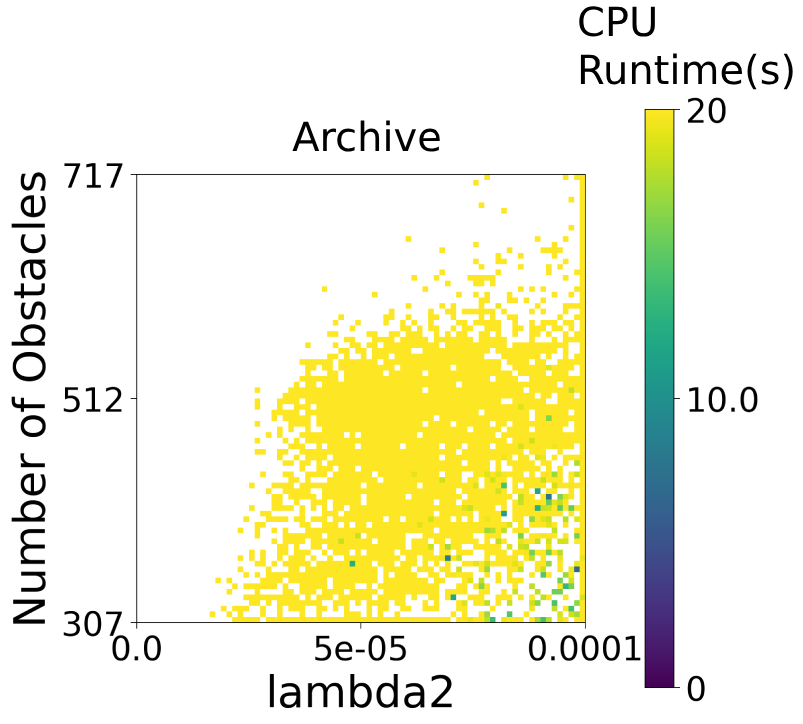}
    \caption{Archive of PBS with the number of obstacles and $\lambda_2$ as diversity measures.}
    \Description{Archive of PBS with the number of obstacles and $\lambda_2$ as diversity measures.}
    \label{fig:obs_lambda2}
\end{figure}

\begin{figure}[!t]
    \centering
    \includegraphics[width=0.4\textwidth]{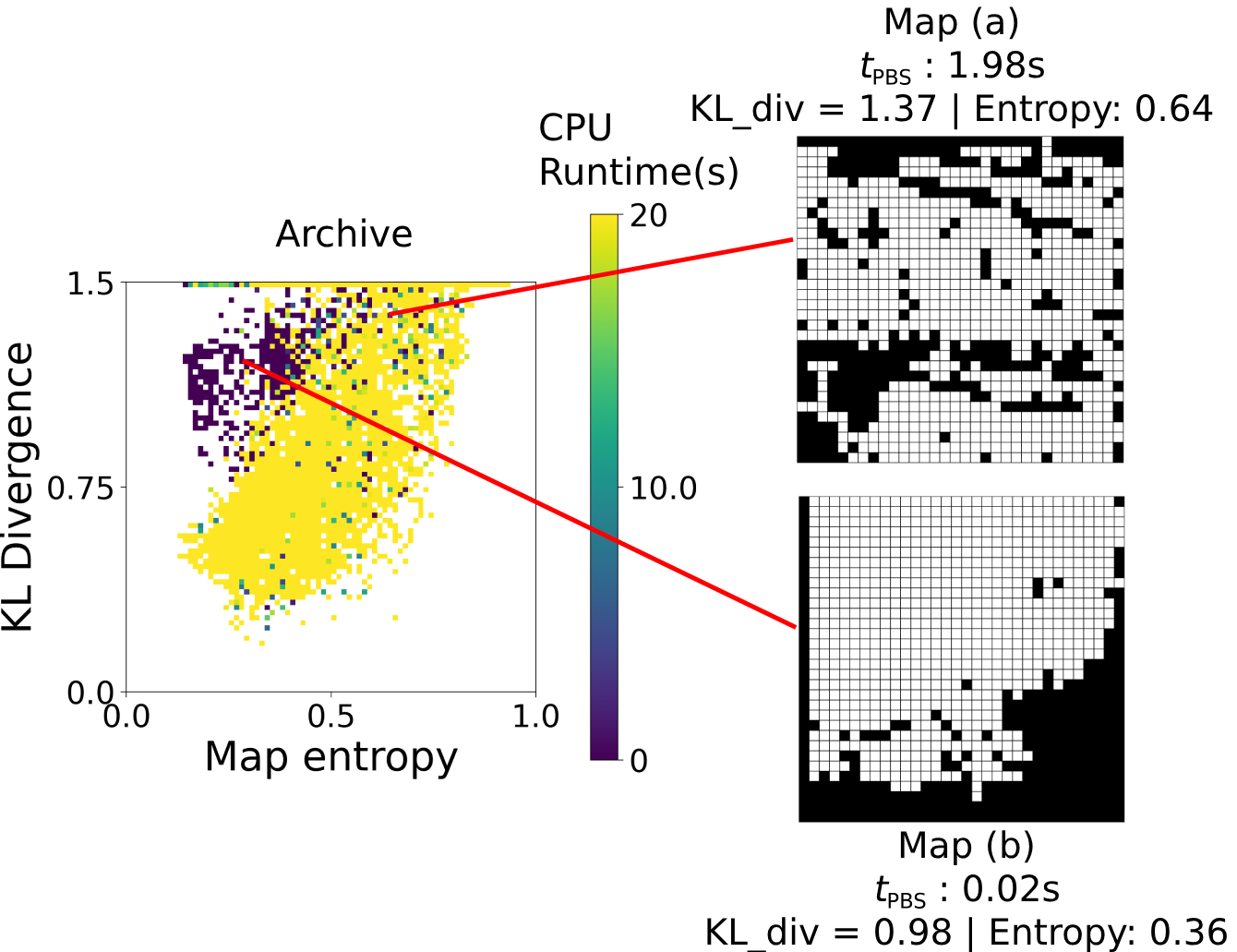}
    \caption{Archive of PBS with map entropy and KL divergence of the tile pattern distribution as diversity measures.}
    \Description{Archive of PBS with map entropy and KL divergence of the tile pattern distribution as diversity measures.}
    \label{fig:entropy}
\end{figure}

\mysubsubsection{Comparison Experiments: Number of obstacles vs. $\boldsymbol{\lambda_2}$.}
We run the one-algorithm experiment of PBS with the number of obstacles and $\lambda_2$ as the diversity measures to explore the relationship between the two measures and their correlation with the hardness of the generated maps. 
From \Cref{fig:obs_lambda2}, we observe that most of the generated maps are hard for PBS to solve. In this case, the QD search is close to convergence, revealing that applying each diversity measure can efficiently generate challenging maps for PBS. Therefore, both diversity measures can quantify the general hardness of generated maps.
We use the number of obstacles in our experiments due to its computational efficiency and intuitive interpretation.


\subsection{Spatial Arrangement} \label{appen:diversity:spatial}

\begin{figure}[!t]
    \centering
    \includegraphics[width=0.4\textwidth]{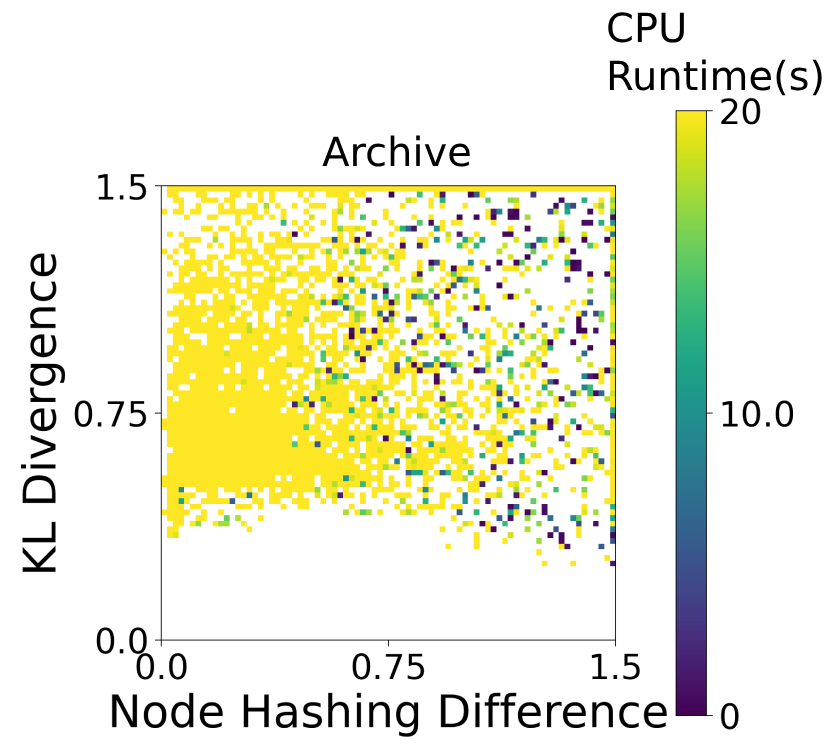}
    \caption{Archive of PBS with KL divergence of the tile pattern distribution and KL divergence of WL graph feature as diversity measure.}
    \Description{Archive of PBS with KL divergence of the tile pattern distribution and KL divergence of WL graph feature as diversity measure.}
    \label{fig:KL_node}
\end{figure}


\mysubsubsection{Entropy of the tile pattern distribution.} 
The entropy of the tile pattern distribution~\cite{ZhangNCA2023} is a diversity measure to indicate the level of pattern regularization. The entropy is calculated as the entropy of the tile pattern distribution mentioned in the KL divergence of the tile pattern distribution section. The lower the entropy, the more regularized patterns the map possesses. 

\mysubsubsection{KL divergence of the Weisfeiler-Lehman (WL) graph feature.} 
We also consider the KL divergence of the Weisfeiler-Lehman (WL) graph feature~\cite{Nino2011WLkernal} between the generated map and the maze maps. The WL graph feature indicates the similarity between graphs. 
To compute the WL graph feature of a map, we view the map as a 4-neighbor grid graph. 
The WL kernel iteratively relabels the nodes based on their neighboring nodes. By examining the distribution of common labels across the relabeling process, the WL kernel captures the local topological features of the graph. We then compute the KL divergence between the graph features of the generated map and the maze maps selected from the MAPF benchmark.
Similar to the tile pattern distribution, a smaller KL divergence of the WL graph feature implies more similarities between the local patterns of the generated map and the maze maps.

\mysubsubsection{Comparison Experiments: KL Divergence of Tile Pattern Distribution vs. Map Entropy.}
We run a one-algorithm experiment with PBS. We use the same objective in \Cref{sec:one-algo} and use map entropy and KL divergence of the tile pattern distribution as the diversity measures. \Cref{fig:entropy} shows the resulting archive of maps. We observe that, with map entropy less than 0.5, the generated maps are usually composed of a large chunk of empty spaces with a large component of obstacles on the boundary, revealing a simple structure.
Compared to maps generated by the KL divergence of the tile distribution in \Cref{fig:PBS} of \Cref{sec:one-algo}, maps generated by map entropy, such as Map (a) and (b), have less diverse patterns. Therefore, we use the KL divergence of tile pattern distribution as the diversity measure.

\begin{figure}[!t]
    \centering
    \includegraphics[width=0.4\textwidth]{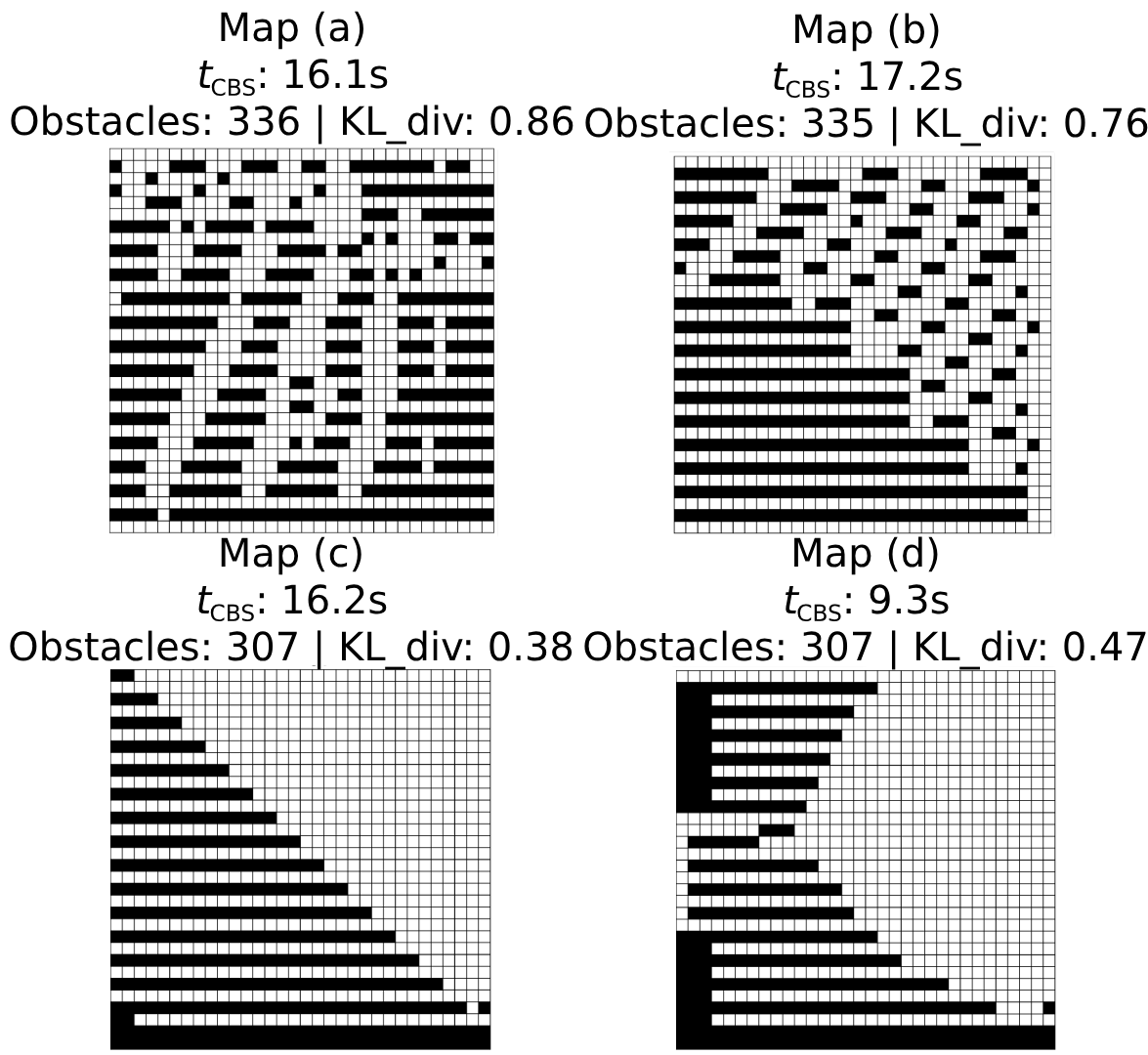}
    \caption{Similar maps with evenly distributed short corridor components or large chunks of empty space on which CBS can solve. Maps (a) and (b) are maps with evenly distributed short corridor components. Map (c) and Map (d) are maps with large chunks of empty space on the right, but also long corridors on the left.}
    \Description{Similar maps with evenly distributed short corridor components or large chunks of empty space on which CBS can solve. Maps (a) and (b) are maps with evenly distributed short corridor components. Map (c) and Map (d) are maps with large chunks of empty space on the right, but also long corridors on the left.}
    \label{fig:cbs_good}
\end{figure}

\begin{figure}[!t]
    \centering
    \includegraphics[width=0.4\textwidth]{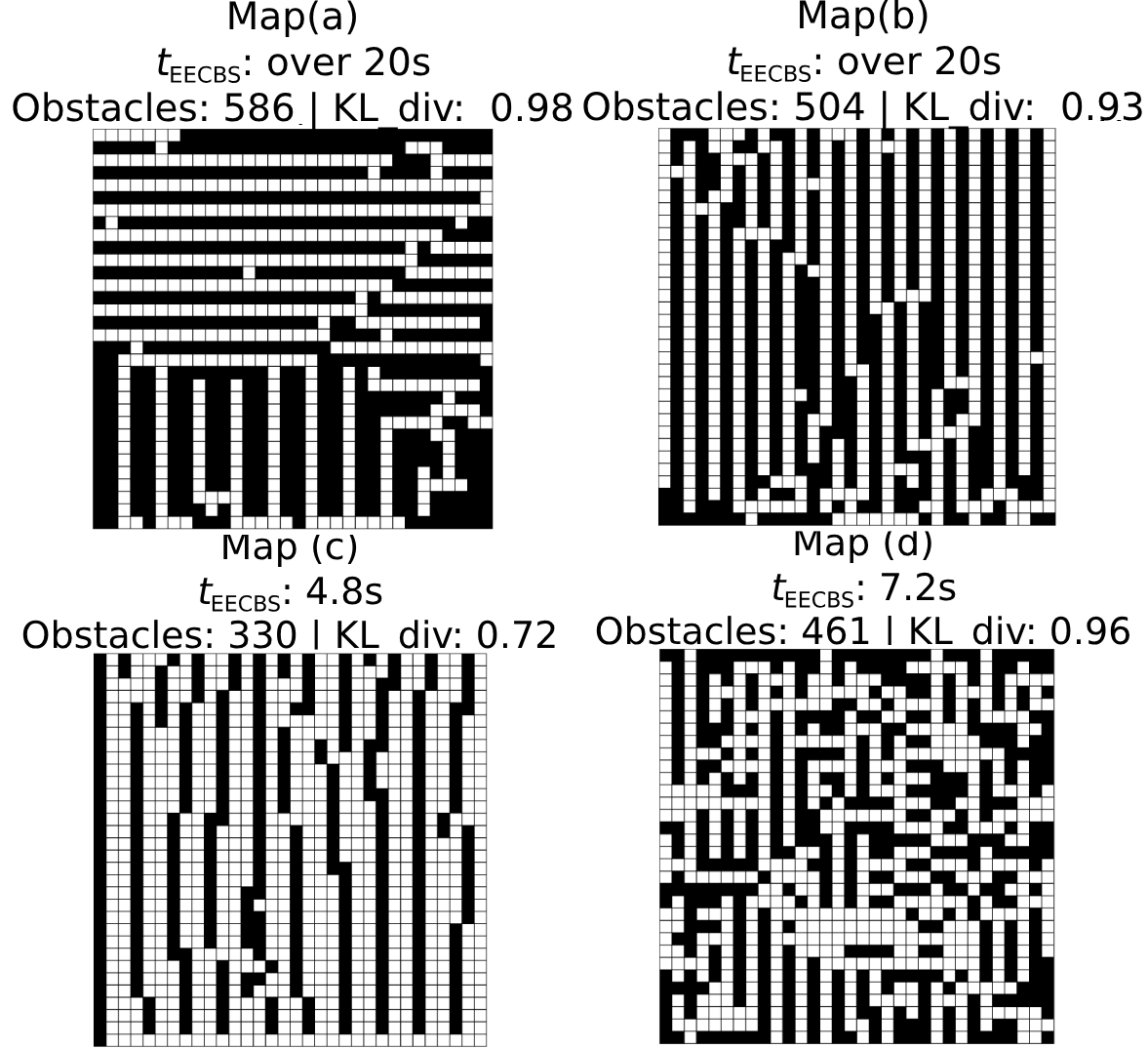  }
    \caption{Map (a) and Map (b) are maps similar to representative maps with long corridors and one-tile entries on which EECBS performs poorly. Map (c) and Map (d) are maps similar to representative maps with two or more columns of empty space between each long obstacle component and short obstacle components on which EECBS performs well.}
    \Description{Map (a) and Map (b) are maps similar to representative maps with long corridors and one-tile entries on which EECBS performs poorly. Map (c) and Map (d) are maps similar to representative maps with two or more columns of empty space between each long obstacle component and short obstacle components on which EECBS performs well.}
    \label{fig:eecbs_bad_good}
\end{figure}

\begin{figure}[!t]
    \centering
    \includegraphics[width=0.4\textwidth]{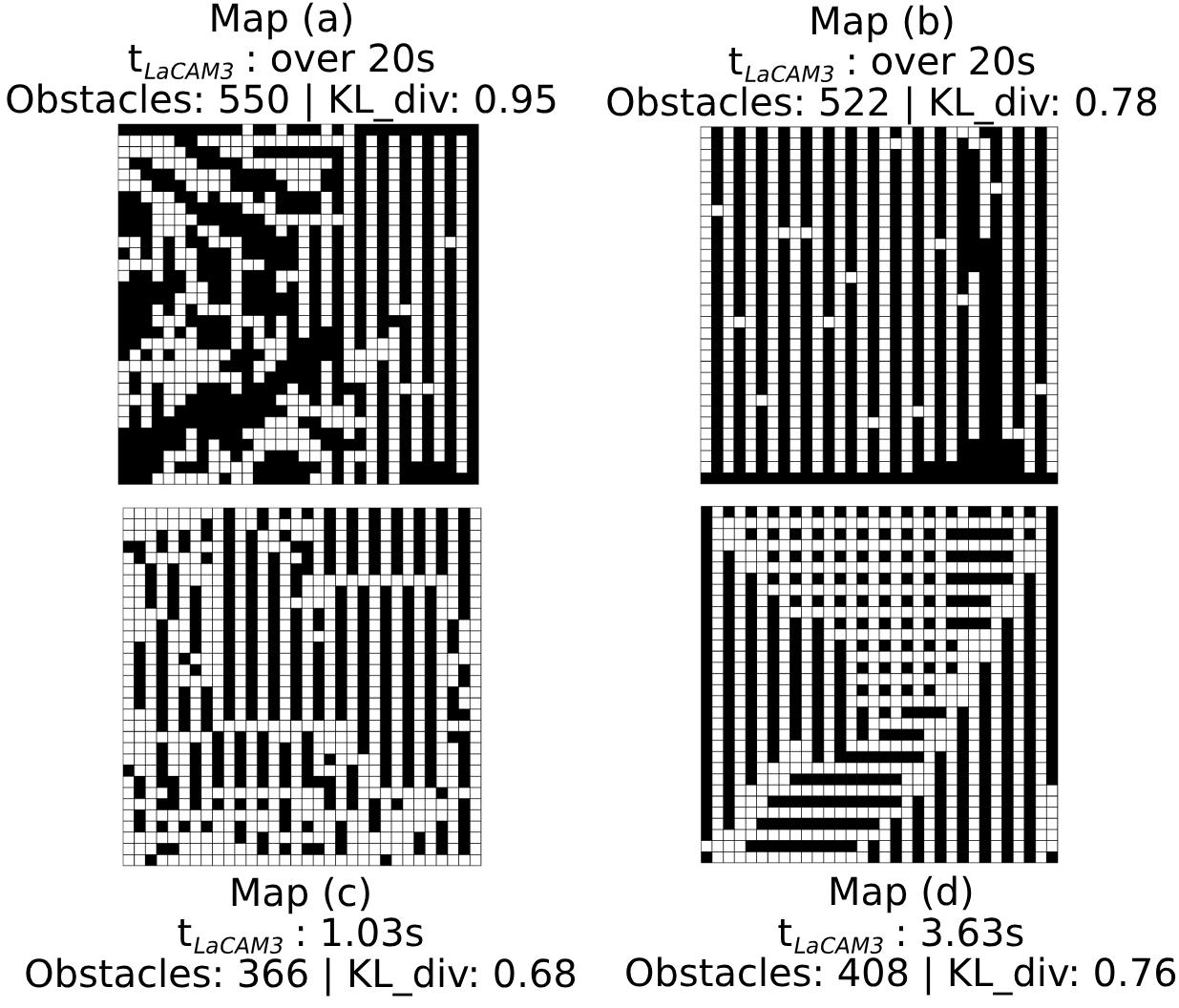}
    \caption{Map (a) and Map (b) are maps similar to representative maps with one-tile entries between corridors on which LaCAM3 performs poorly. Map (c) and Map (d) are maps similar to representative maps of more empty spaces between long obstacle components and short obstacle components on which LaCAM3 performs well.}
    \Description{Map (a) and Map (b) are maps similar to representative maps with one-tile entries between corridors on which LaCAM3 performs poorly. Map (c) and Map (d) are maps similar to representative maps of more empty spaces between long obstacle components and short obstacle components on which LaCAM3 performs well.}
    \label{fig:LaCAM3_bad_good}
\end{figure}

\begin{figure}[!t]
    \centering
    \includegraphics[width=0.4\textwidth]{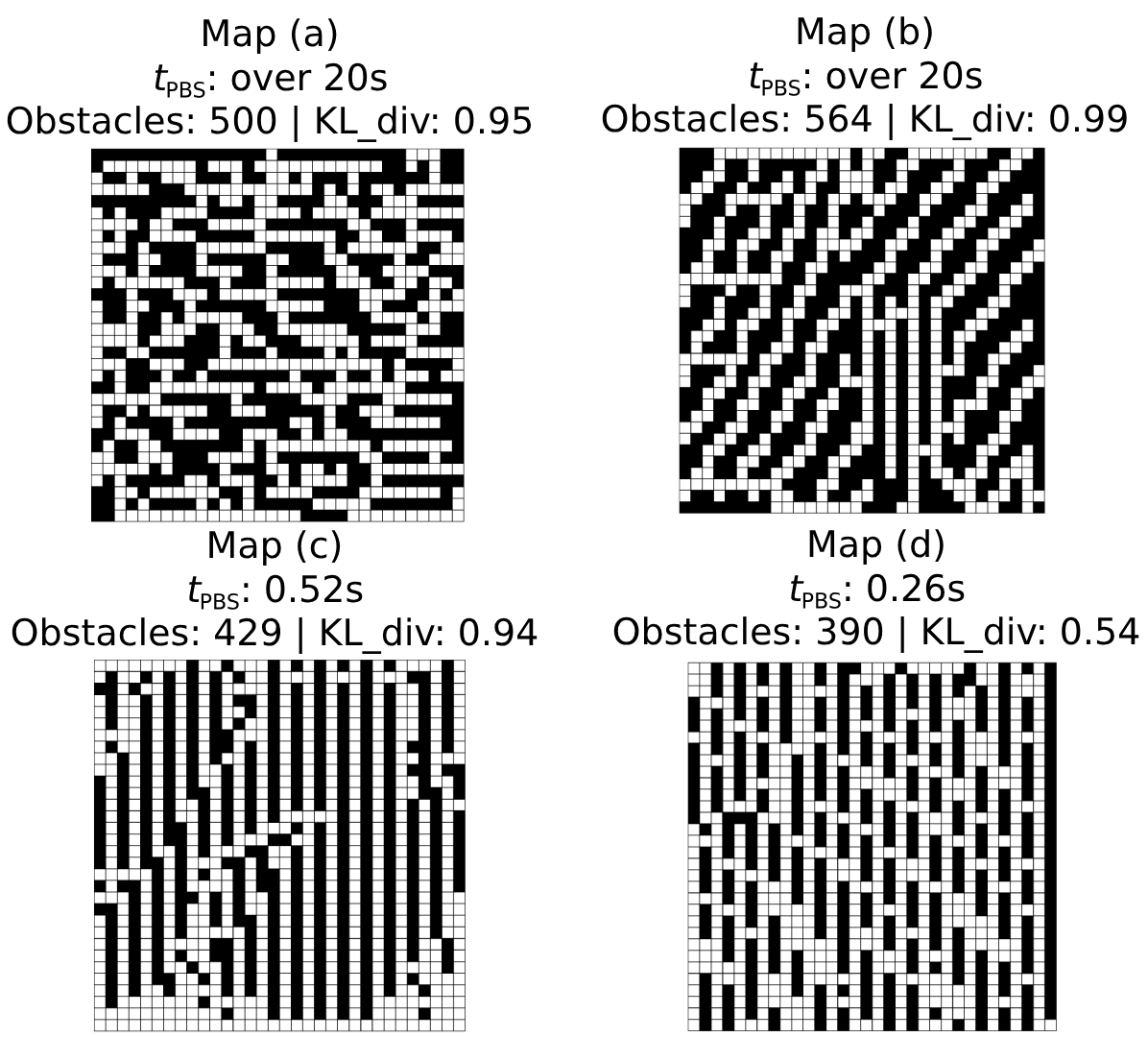}
    \caption{Map (a) and Map (b) are maps similar to representative maps with one-tile entries between corridors on which PBS performs poorly. Map (c) and Map (d) are maps similar to representative maps of long corridors with more entry spaces and short obstacle components on which PBS performs well.}
    \Description{Map (a) and Map (b) are maps similar to representative maps with one-tile entries between corridors on which PBS performs poorly. Map (c) and Map (d) are maps similar to representative maps of long corridors with more entry spaces and short obstacle components on which PBS performs well.}
    \label{fig:pbs_bad_good}
\end{figure}

\begin{figure}[!t]
    \centering
    \includegraphics[width=0.4\textwidth]{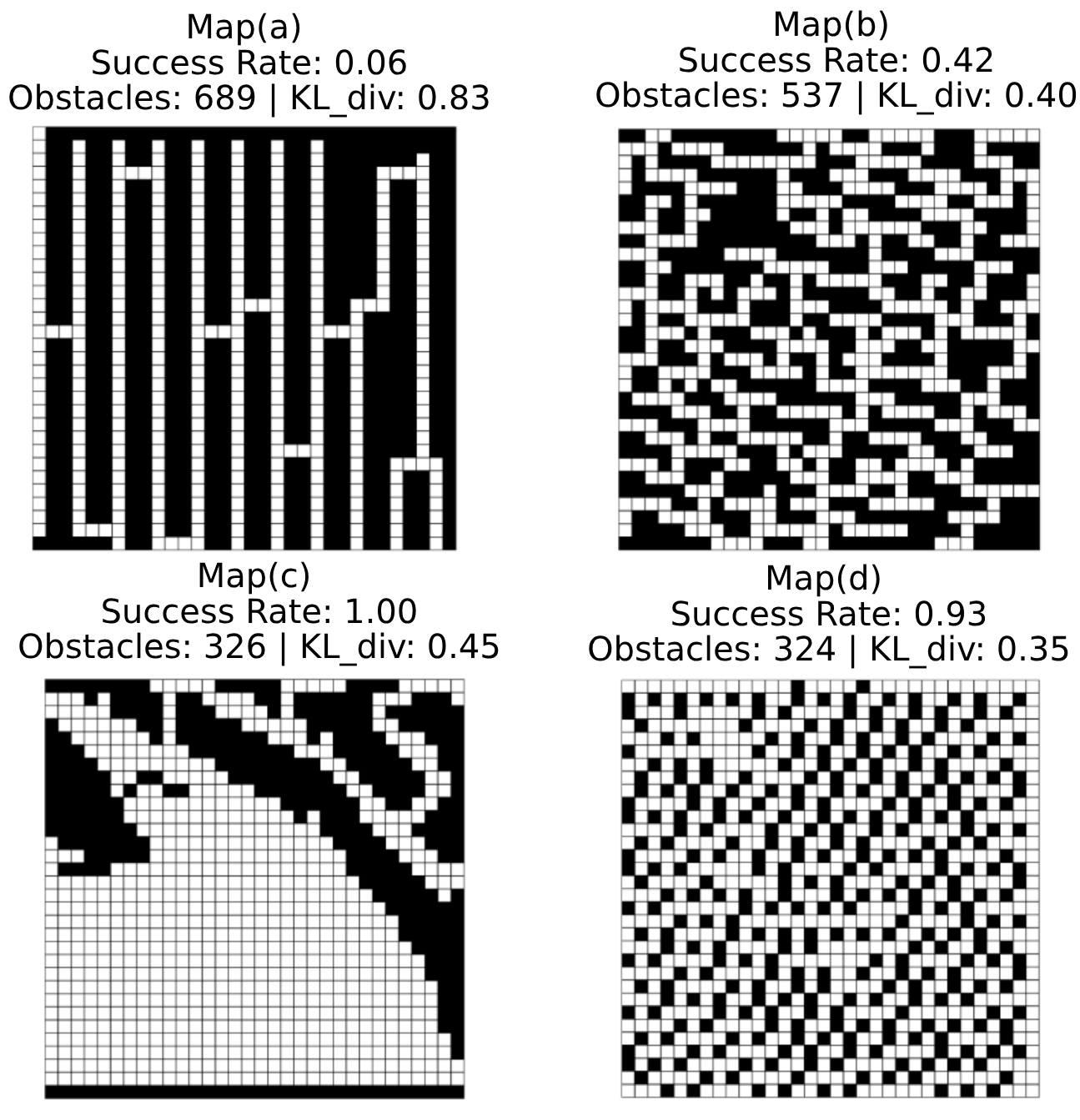}
    \caption{Map (a) and Map (b) are maps similar to representative maps with long corridors and one-entry spaces with many obstacles on which PIBT performs poorly. Map (c) and Map (d) are maps similar to representative maps with large chunks of empty space and one-entry spaces with fewer obstacles on which PIBT performs well.}
    \Description{Map (a) and Map (b) are maps similar to representative maps with long corridors and one-entry spaces with many obstacles on which PIBT performs poorly. Map (c) and Map (d) are maps similar to representative maps with large chunks of empty space and one-entry spaces with fewer obstacles on which PIBT performs well.}
    \label{fig:pibt_bad_good}
\end{figure}

\begin{figure}[!t]
    \centering
    \includegraphics[width=0.4\textwidth]{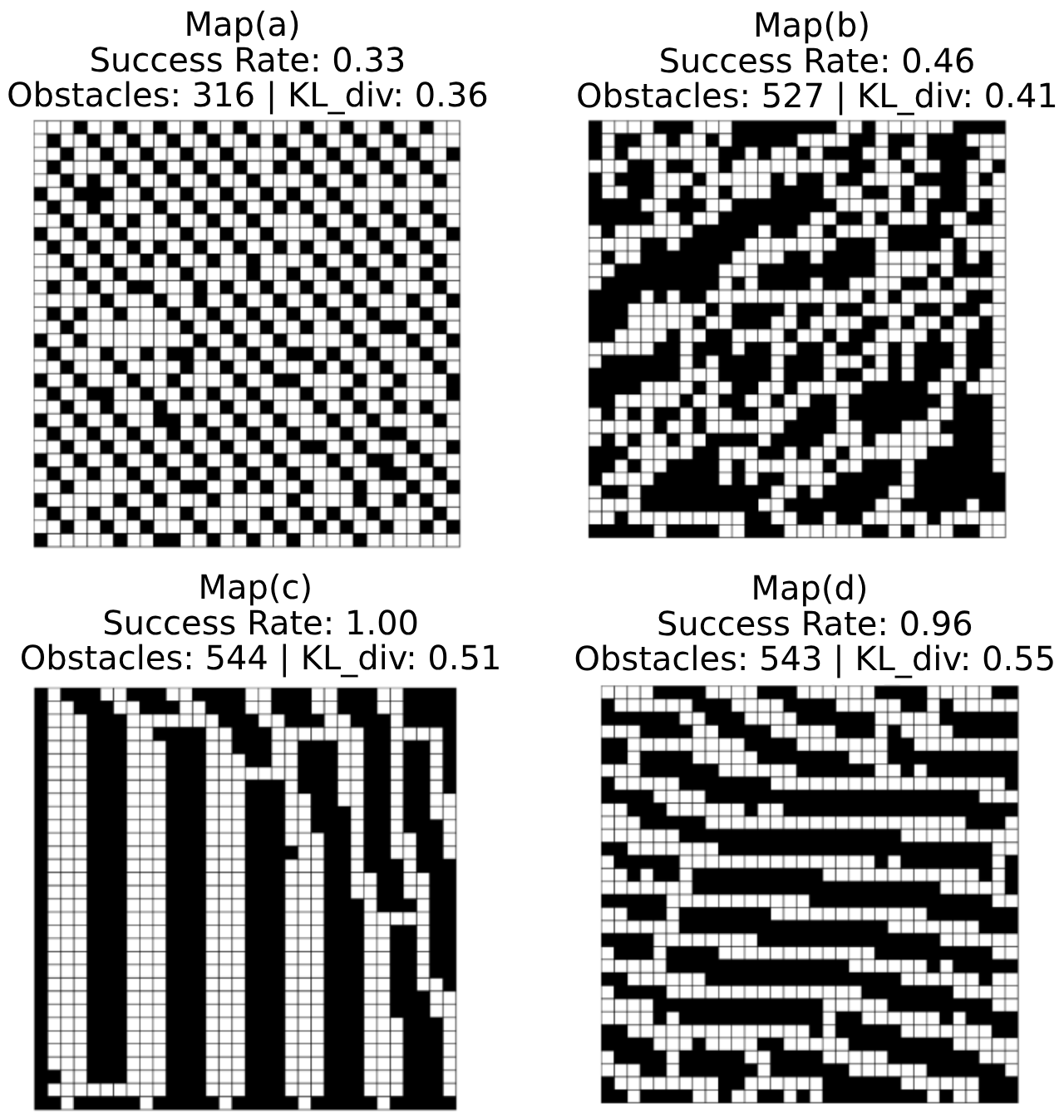}
    \caption{Map (a) and Map (b) are maps similar to representative maps with long corridors and one-entry spaces on which LTF performs poorly. Map (c) and Map (d) are maps similar to representative maps with fewer obstacles on which LTF performs well. }
    \Description{Map (a) and Map (b) are maps similar to representative maps with long corridors and one-entry spaces on which LTF performs poorly. Map (c) and Map (d) are maps similar to representative maps with fewer obstacles on which LTF performs well. }
    \label{fig:LTF_bad_good}
\end{figure}

\begin{figure}[!t]
    \centering
    \includegraphics[width=0.45\textwidth]{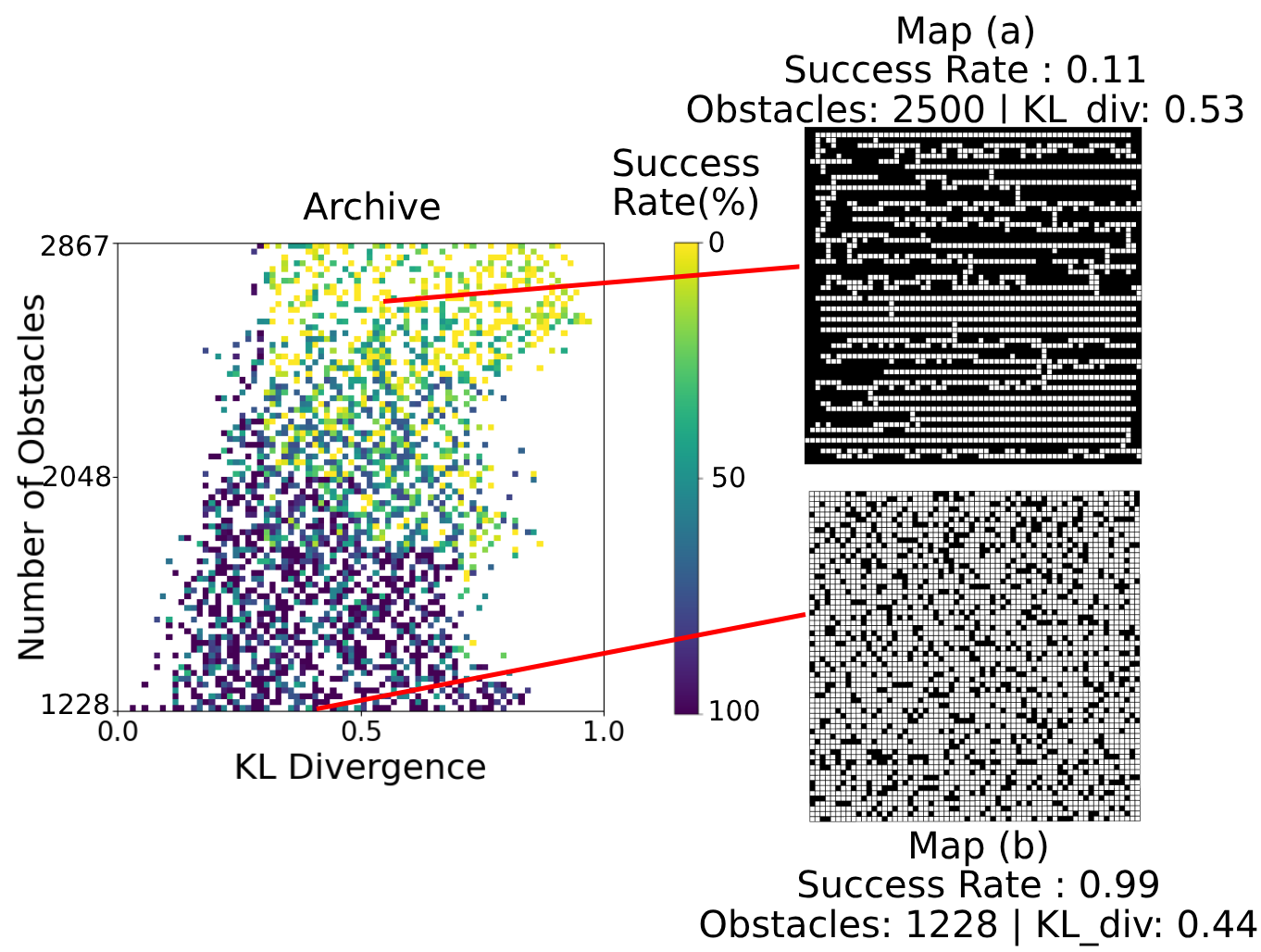}
    \caption{Archive for PIBT with representative maps of size 64 $\times$ 64.} 
    \Description{Archive for PIBT with representative maps of size 64 $\times$ 64.}
    \label{fig:PIBT_64}
\end{figure}

\begin{table}
    \centering
        \caption{
    Validation experiments results of generated maps in one-algorithm experiments. We run 200 MAPF instances in the generated maps and report the success rate and average runtime.
    }
    \label{tab:200-validation}
    \small
        \begin{tabular}{c|cccc}
        \toprule
        MAPF Algos & Map & Runtime & Success Rate\\
        \midrule
        \multirow{4}{*}{CBS (\cref{fig:CBS})} & Map (a) & over 20s & $0\%$  \\
        & Map (b) & 6.1s & $78\%$  \\
        & Map (c) & over 20s & $0\%$  \\
         & Map (d) & over 20s & $0\%$  \\
        \midrule
        \multirow{4}{*}{EECBS (\cref{fig:EECBS})}
        & Map (a) & over 2.5s & $97\%$  \\
        & Map (b) & 18.3s & $5\%$  \\
         & Map (c) & over 20s & $0\%$  \\
         & Map (d) & 0.08s & $100\%$  \\
        \midrule
        \multirow{4}{*}{PBS (\cref{fig:PBS})} 
        & Map (a) & 3.4s & $87\%$  \\
        & Map (b) & 3.6s & $88\%$  \\
         & Map (c) & No solution & $0\%$  \\
         & Map (d) & 19.1s & $17\%$  \\
        \midrule
         \multirow{4}{*}{LaCAM3 (\cref{fig:LaCAM3})} 
        & Map (a) & 19.5s & $11\%$  \\
        & Map (b) & 19.3s & $5\%$  \\
         & Map (c) & 16.5s & $18\%$  \\
         & Map (d) & 2.1s & $98\%$  \\
        \midrule
        \multirow{4}{*}{PIBT (\cref{fig:PIBT})} & Map (a) & \textbackslash & $96\%$  \\
         & Map (b) & \textbackslash & $95\%$  \\
         & Map (c) & \textbackslash & $12\%$  \\
         & Map (d) & \textbackslash & $53\%$  \\
        \midrule
        \multirow{4}{*}{LTF (\cref{fig:LTF})} 
        & Map (a) & \textbackslash & $8\%$  \\
        & Map (b) & \textbackslash & $27\%$  \\
        & Map (c) & \textbackslash & $98\%$  \\
        & Map (d) & \textbackslash & $30\%$  \\
        \bottomrule
        \end{tabular}
\end{table}

\mysubsubsection{Comparison Experiments: KL Divergence of Tile Pattern Distribution vs. KL Divergence of WL Graph Feature.}
We run the one-algorithm experiment of PBS with diversity measures being the KL divergence of the tile pattern distribution and the KL divergence of the WL graph feature.
From \Cref{fig:KL_node}, we observe that all generated maps are evenly distributed across the archive, revealing a similar distribution of generated maps with different pairs of KL divergence of the tile pattern distribution and the KL divergence of WL graph feature.
Therefore, these two diversity measures are similar. Both quantify the similarity between the generated maps and the pre-defined set of maze maps~\cite{SternSoCS19}. 
We use the KL divergence of the tile pattern distribution in our experiments since the WL graph feature needs a graph transformation and several iterations of hash transformation, which are more computationally expensive than the tile pattern distribution.



\section{Computational Cost and Compute Resource} \label{appen:implement-compute}

\mysubsubsection{Computational Cost.}
QD-MAPPER relies on QD algorithms, which require a massive number of evaluations in the generated maps. Specifically, with our one-algorithm experiment setup, it takes 6 to 7 hours to generate diverse maps for PIBT and LTF, 10 hours for PBS, and more than 12 hours for CBS and EECBS. 
Future works can focus on reducing the computational cost of QD algorithms by using surrogate models~\cite{Zhang2021DeepSA}.
Meanwhile, we will release our most representative generated set of diverse maps online. If the researchers' new algorithm is built on one of the algorithms we have tested, they can directly evaluate new algorithms on our generated set of diverse maps.


\mysubsubsection{Compute Resources.}
We run our experiments on three machines: (1) a local machine equipped with a 64-core AMD Ryzen Threadripper 3990X CPU and 192 GB of RAM, (2) a local machine with a 64-core AMD Ryzen Threadripper 7980X CPU and 128 GB of RAM, and (3) a high-performance cluster~\cite{PSCBridgeTwo2021} featuring multiple 64-core AMD EPYC 7742 CPUs, each with 256 GB of RAM. All CPU runtimes are measured on machine (1).

\section{Maps with similar patterns} \label{appen:similar_maps}

\Cref{fig:cbs_good} shows the easy maps of CBS, similar to the maps shown in \Cref{fig:CBS} in \Cref{sec:one-algo} in the main text.
\Cref{fig:eecbs_bad_good} shows the challenging and easy maps of EECBS, similar to the maps shown in \Cref{fig:EECBS} in \Cref{sec:one-algo} in the main text.
\Cref{fig:LaCAM3_bad_good} shows the challenging and easy maps of LaCAM3, similar to the maps shown in \Cref{fig:LaCAM3} in \Cref{sec:one-algo} in the main text.
\Cref{fig:pbs_bad_good} shows the challenging and easy maps of PBS, similar to the maps shown in \Cref{fig:PBS} in \Cref{sec:one-algo} in the main text.
\Cref{fig:pibt_bad_good} shows the challenging and easy maps of PIBT, similar to the maps shown in \Cref{fig:PIBT} in \Cref{sec:one-algo} in the main text.
\Cref{fig:LTF_bad_good} shows the challenging and easy maps of LTF, similar to the maps shown in \Cref{fig:LTF} in \Cref{sec:one-algo} in the main text.



\section{Validation Results} \label{appen:validation}

\Cref{tab:200-validation} shows the validation results of one-algorithm experiments in \Cref{sec:one-algo}. 

\section{Archive of PIBT with generated larger maps} \label{appen:64}

\Cref{fig:PIBT_64} shows the archive of PIBT by generating maps of size 64 
$\times$ 64. 


\end{document}